\documentclass[compsoc]{IEEEtran}
\IEEEoverridecommandlockouts

\PassOptionsToPackage{dvipsnames}{xcolor}
\PassOptionsToPackage{switch, pagewise}{lineno}

\usepackage[newfloat,cachedir=.]{minted}
\usepackage[T1]{fontenc}
\usepackage[inline]{enumitem}
\usepackage[pdftex]{graphicx}
\usepackage[table,xcdraw]{xcolor}
\graphicspath{figures}
\DeclareGraphicsExtensions{.pdf}
\usepackage{amsmath,bbm}
\usepackage{algorithm}
\usepackage{boldline}
\usepackage[noend]{algpseudocode}
\usepackage{array}
\usepackage{adjustbox}
\usepackage{bigstrut}
\usepackage{stfloats}
\usepackage{soul}
\usepackage{url}
\usepackage{fontawesome}
\usepackage{xspace}
\usepackage{colortbl}
\usepackage{boldline}
\usepackage{hyperref}
\usepackage[capitalise]{cleveref}
\usepackage[framemethod=tikz]{mdframed}
\usepackage[scale=0.90]{plex-mono}
\usepackage{multirow}
\usepackage{lettrine}
\usepackage{tikz}
\usepackage{upquote}
\usepackage{nicefrac}
\usetikzlibrary{fit, tikzmark}
\usepackage{textcomp}
\usepackage{enumitem}
\setlist{nosep}
\usepackage{rotating}
\usepackage{booktabs}
\usepackage{multicol}
\usepackage{blkarray}
\usepackage{wrapfig}
\usepackage{graphics}
\usepackage{adjustbox}
\usepackage{tabularray}
\usepackage{makecell}
\usepackage{listings, listings-rust, listings-algo}
\usepackage{tablefootnote}
\usepackage{float}
\usepackage[numbers, sort]{natbib}
\usepackage{subcaption}

\newcommand{\bi}{\begin{itemize}[leftmargin=*, wide=0pt]}
\newcommand{\ei}{\end{itemize}}
\newcommand{\beq}{\begin{equation}}
\newcommand{\eeq}{\end{equation}}
\newcommand{\be}{\begin{enumerate}[leftmargin=*, wide=0pt]}
\newcommand{\ee}{\end{enumerate}}

\def\rot{\rotatebox}
\newcommand{\tool}{\textsc{C2SaferRust}\xspace}

\newcommand{\rebuttal}[2]{%
    #2\unskip
}

\newcommand{\revision}[1]{%
    #1\unskip
}

\newmdenv[
    tikzsetting= {fill=blue!12},
    skipabove=0.5em,
    skipbelow=0.5em,
    linewidth=2pt,
    innerleftmargin=2pt,
    innerrightmargin=2pt,
    innertopmargin=2pt,
    innerbottommargin=2pt,
    linecolor=gray80,
    roundcorner=2pt, 
    shadow=true,
    shadowsize=5pt,
    shadowcolor=gray60
]{myshadowbox}
\newenvironment{result}
{\begin{myshadowbox} \textbf{Summary:}\xspace}
{\end{myshadowbox}}

\definecolor{bluekeywords}{rgb}{0,0,1}
\definecolor{greencomments}{rgb}{0,0.5,0}
\definecolor{greennumbers}{rgb}{0,0.5,0}
\definecolor{redstrings}{rgb}{0.64,0.08,0.08}
\definecolor{blueident}{rgb}{0, 0.08, 0.45}

\lstdefinestyle{myc}{
  language=c,
  showstringspaces=false,
  basicstyle=\small\ttfamily,
  keywordstyle=\color{bluekeywords},
  commentstyle=\color{greencomments},
  identifierstyle=\color{blueident},
  stringstyle=\color{redstrings},
  frame=none,
  backgroundcolor=\color{white},
  escapeinside={<@}{@>},
  columns=fullflexible,
  breakatwhitespace=true,
  breaklines=true,
  literate=
  *{0}{{{\color{greennumbers}0}}}1
  {1}{{{\color{greennumbers}1}}}1
  {2}{{{\color{greennumbers}2}}}1
  {3}{{{\color{greennumbers}3}}}1
  {4}{{{\color{greennumbers}4}}}1
  {5}{{{\color{greennumbers}5}}}1
  {6}{{{\color{greennumbers}6}}}1
  {7}{{{\color{greennumbers}7}}}1
  {8}{{{\color{greennumbers}8}}}1
  {9}{{{\color{greennumbers}9}}}1
}

\newcommand{\eg}{\hbox{\emph{e.g.,}}\xspace}
\newcommand{\ie}{\hbox{\emph{i.e.,}}\xspace}

\definecolor{javared}{rgb}{0.6,0,0} 
\definecolor{javagreen}{rgb}{0.25,0.5,0.35} 
\definecolor{javapurple}{rgb}{0.5,0,0.35} 
\definecolor{javadocblue}{rgb}{0.25,0.35,0.75} 

\definecolor{beforeColor}{RGB}{255,153,153} 
\definecolor{afterColor}{RGB}{102,102,153}  

\newcommand{\TT}[1]{{\javadocblue{\texttt{\small#1}}}}

\newcommand*\circled[1]{\tikz[baseline=(char.base)]{
            \node[shape=circle,fill=gray,inner sep=2pt] (char) {\footnotesize\textcolor{white}{\bfseries #1}};}}

\definecolor{bluekeywords}{rgb}{0,0,1}
\definecolor{greencomments}{rgb}{0,0.5,0}
\definecolor{redstrings}{rgb}{0.64,0.08,0.08}
\definecolor{xmlcomments}{rgb}{0.5,0.5,0.5}
\definecolor{types}{rgb}{0.17,0.57,0.68}

\crefformat{footnote}{#2\footnotemark[#1]#3}

\algdef{SE}[DOWHILE]{Do}{doWhile}{\algorithmicdo}[1]{\algorithmicwhile\ #1}%

\let\origthelstnumber\thelstnumber
\makeatletter
\newcommand*\Suppressnumber{%
  \lst@AddToHook{OnNewLine}{%
    \let\thelstnumber\relax%
     \advance\c@lstnumber-\@ne\relax%
    }%
}

\newcommand*\Reactivatenumber[1]{%
  \setcounter{lstnumber}{\numexpr#1-1\relax}
  \lst@AddToHook{OnNewLine}{%
   \let\thelstnumber\origthelstnumber%
   \refstepcounter{lstnumber}
  }%
}

\makeatletter
\newcommand{\crefnames}[3]{%
  \@for\next:=#1\do{%
    \expandafter\crefname\expandafter{\next}{#2}{#3}%
  }%
}
\makeatother
\crefnames{part,chapter,section}{\S}{\S\S}

\newcommand\javadocblue[1]{\textcolor[rgb]{0.25,0.35,0.75}{{#1}}}

\definecolor{blueish}{RGB}{250, 250, 255}
\definecolor{greenish}{RGB}{250, 255, 250}
\definecolor{redish}{RGB}{255, 200, 200}

\definecolor{highlight}{RGB}{175, 255, 100}
\definecolor{gray01}{gray}{.98}
\definecolor{gray05}{gray}{0.95}
\definecolor{gray08}{gray}{0.92}
\definecolor{gray10}{gray}{0.90}
\definecolor{gray12}{gray}{0.88}
\definecolor{gray15}{gray}{0.85}
\definecolor{gray18}{gray}{0.82}
\definecolor{gray20}{gray}{0.80}
\definecolor{gray25}{gray}{0.75}
\definecolor{gray30}{gray}{0.70}
\definecolor{gray35}{gray}{0.65}
\definecolor{gray40}{gray}{0.60}
\definecolor{gray45}{gray}{0.55}
\definecolor{gray50}{gray}{0.50}
\definecolor{gray55}{gray}{0.45}
\definecolor{gray60}{gray}{0.40}
\definecolor{gray65}{gray}{0.35}
\definecolor{gray70}{gray}{0.30}
\definecolor{gray75}{gray}{0.25}
\definecolor{gray80}{gray}{0.20}
\definecolor{gray85}{gray}{0.15}
\definecolor{gray90}{gray}{0.10}
\definecolor{gray95}{gray}{0.05}

\definecolor{gray05}{gray}{0.95}
\definecolor{gray08}{gray}{0.92}
\definecolor{gray10}{gray}{0.90}
\definecolor{gray12}{gray}{0.88}
\definecolor{gray15}{gray}{0.85}
\definecolor{gray18}{gray}{0.82}
\definecolor{gray20}{gray}{0.80}
\definecolor{gray25}{gray}{0.75}
\definecolor{gray30}{gray}{0.70}
\definecolor{gray35}{gray}{0.65}
\definecolor{gray40}{gray}{0.60}
\definecolor{gray45}{gray}{0.55}
\definecolor{gray50}{gray}{0.50}
\definecolor{gray55}{gray}{0.45}
\definecolor{gray60}{gray}{0.40}
\definecolor{gray65}{gray}{0.35}
\definecolor{gray70}{gray}{0.30}
\definecolor{gray75}{gray}{0.25}
\definecolor{gray80}{gray}{0.20}
\definecolor{gray85}{gray}{0.15}
\definecolor{gray90}{gray}{0.10}
\definecolor{gray95}{gray}{0.05}
\definecolor{blue}{HTML}{0f62fe}
\definecolor{magenta}{HTML}{d02670}
\definecolor{red}{RGB}{234,51,35}
\definecolor{ibm-blue-light}{HTML}{a6c8ff}

\hypersetup{
    colorlinks=true,
    linkcolor=blue,
    filecolor=magenta,      
    urlcolor=blue,
    citecolor=blue,
    pdftitle={Testing The Waters},
    pdfpagemode=FullScreen,
    }
\setminted{
    breaklines,
    frame=lines,
    bgcolor=white,
    framesep=1mm,
    baselinestretch=1.0,
    fontsize=\scriptsize, 
    breaklines, 
    breakanywhere, 
    linenos,
    numbersep=2pt,
    mathescape=true,
    style=xcode,
    escapeinside=||
}

\begin{document}
\usetikzlibrary{shadows}
\title{\tool: Transforming C Projects into Safer Rust with NeuroSymbolic Techniques}


\author{Vikram Nitin,~%
    Rahul Krishna,~%
    Luiz Lemos do Valle,~\and%
    Baishakhi Ray

\IEEEcompsocitemizethanks{%
\IEEEcompsocthanksitem Vikram Nitin, Luiz Lemos do Valle and Baishakhi Ray are with the Department
of Computer Science, Columbia University, New York, NY.
E-mail: \{\href{mailto:vikram.nitin@columbia.edu}{vikram.nitin}, \href{mailto:lld2131@columbia.edu}{lld2131}\}@columbia.edu, \href{mailto:rayb@cs.columbia.edu}{rayb@cs.columbia.edu}.
\IEEEcompsocthanksitem Rahul Krishna is with IBM T.J. Watson Research Center, Yorktown Heights, NY.
E-mail: \href{mailto:rkrsn@ibm.com}{rkrsn@ibm.com}.
}
}

\IEEEtitleabstractindextext{%
\begin{abstract}
In recent years, there has been a lot of interest in converting C code to Rust, to benefit from the memory and thread safety guarantees of Rust. C2Rust is a rule-based system that can automatically convert C code to functionally identical Rust, but the Rust code that it produces is non-idiomatic, i.e., makes extensive use of unsafe Rust, a subset of the language that \textit{doesn't} have memory or thread safety guarantees. At the other end of the spectrum are LLMs, which produce idiomatic Rust code, but these have the potential to make mistakes and are constrained in the length of code they can process. In this paper, we present \tool, a novel approach to translate C to Rust that combines the strengths of C2Rust and LLMs. We first use C2Rust to convert C code to non-idiomatic, unsafe Rust. We then decompose the unsafe Rust code into slices that can be individually translated to safer Rust by an LLM. After processing each slice, we run end-to-end test cases to verify that the code still functions as expected. We also contribute a benchmark of 7 real-world programs, translated from C to unsafe Rust using C2Rust. Each of these programs also comes with end-to-end test cases. On this benchmark, we are able to reduce the number of raw pointers by up to 38\%, and reduce the amount of unsafe code by up to 28\%, indicating an increase in safety. The resulting programs still pass all test cases. \tool also shows convincing gains in performance against two previous techniques for making Rust code safer.
\end{abstract}}
\maketitle
\IEEEdisplaynontitleabstractindextext
\IEEEdisplaynontitleabstractindextext
\ifCLASSOPTIONcaptionsoff
 \newpage
\fi

\maketitle

\section{Introduction}

The Rust language has emerged as a popular alternative to C, thanks to its strong memory safety and thread safety guarantees. Rust uses a memory model based on ownership and borrowing, which prevents accesses of a memory location after it has been de-allocated or freed. It also disallows mutable aliasing, which prevents data races in multi-threaded programs. At the same time, Rust has comparable runtime performance to C, making it suitable for performance-critical applications like device drivers and embedded systems.

In contrast, C code can contain both memory safety and thread safety errors, which makes it prone to security vulnerabilities. For all of these reasons, there has been a lot of interest in converting all C code to Rust. However, doing this manually is challenging and requires tremendous manpower and expertise. Just a single project, the Linux kernel, contains more than 7 million lines of C code, spread over more than 16,000 files. Thus, there is a need to develop tools to \textit{automatically} convert C to Rust.

\noindent
\textbf{Limitations of Existing Approaches:} Existing approaches for code translation fall into two broad categories. The first is \textit{transpilers} or cross-compilers, which are rule-based systems. For instance, the C2Rust project \cite{c2rust} is a \textit{transpiler} that converts C code to Rust. Although the Rust code produced by C2Rust is almost always correct, it is \textit{unsafe} Rust, a subset of the language that lacks the memory and thread safety guarantees usually associated with Rust. Further, this code is \textit{non-idiomatic}, \ie, very different from the kind of Rust code usually written by humans, making it hard to read and maintain.

There have been a few approaches \cite{emre2021translating, zhang2023ownership, hong2024don, hong2024tag} that augment the transpiler framework with specific rule-based transformations, such as converting certain datatypes to other types. However, these are limited in scope, and cannot ``zoom out'' to transform the code at a higher granularity.

The second category of approaches is using Large Language Models (LLMs). These models are trained on vast amounts of human-written code, and thus produce code that is \textit{idiomatic}, \ie uses the most conventional idioms and constructs of a programming language. However, this code frequently has errors and is unreliable. This problem is compounded because LLM-generated code has the \textit{appearance} of correctness, which makes any errors harder to spot.

Another challenge when applying LLMs for code translation is the \textit{length} of input and/or output code. LLMs operate on a fixed context window, which means that they can only process inputs and generate outputs that are less than a certain predefined length. A real-world code project with dozens of files and hundreds of lines per file may not fit within this window. Further, even if the entire project does fit, it is known that LLMs tend to produce more errors as the length of the input or output increases \rebuttal{[Reviewer 3 - citation for this claim]}{\cite{peng2023yarn}}. For these reasons, most LLM-based code translation research has focused on programs that are relatively short standalone files, such as those used in competitive programming contests.

\noindent
\textbf{Our Approach:} In this paper, we combine the strengths of the transpiler approach and the LLM approach to design \tool - an algorithm to convert full real-world C projects into Rust. Tackling an entire codebase at once can lead to inefficiencies and increased complexity. Instead, we propose breaking down the translation process into smaller, manageable components, such as individual methods or functions. These smaller units can be translated first, and once translated, the changes can be progressively integrated into the larger project by carefully propagating the altered parameters, variables, and interfaces to the rest of the codebase. This requires strategic planning to ensure that dependencies between components are maintained and that changes flow seamlessly throughout the system.

\tool uses the C2Rust transpiler in the first step to convert all the C code to non-idiomatic \textit{unsafe} Rust. Then, we incrementally convert slices of this code to safe Rust using an LLM. The transpiled unsafe Rust code thus acts as a \textit{bridge} between the original C code and the LLM-generated safe Rust code. We traverse the parse tree of the transpiled code to decompose it into smaller units which can be individually translated.

We use static program analysis to analyze dependencies between these units and arrange them in an optimal order for translation. Then we augment each unit with context information, again derived using static analysis, and prompt an LLM with this information to translate the code into safe Rust. We ensure that each unit is translated correctly by compiling the code and running end-to-end test cases each time an LLM produces a translation. If there are errors, we provide the compiler or test feedback to the LLM and iteratively refine the translations.

\begin{figure*}
    \centering
    \includegraphics[width=0.60\linewidth]{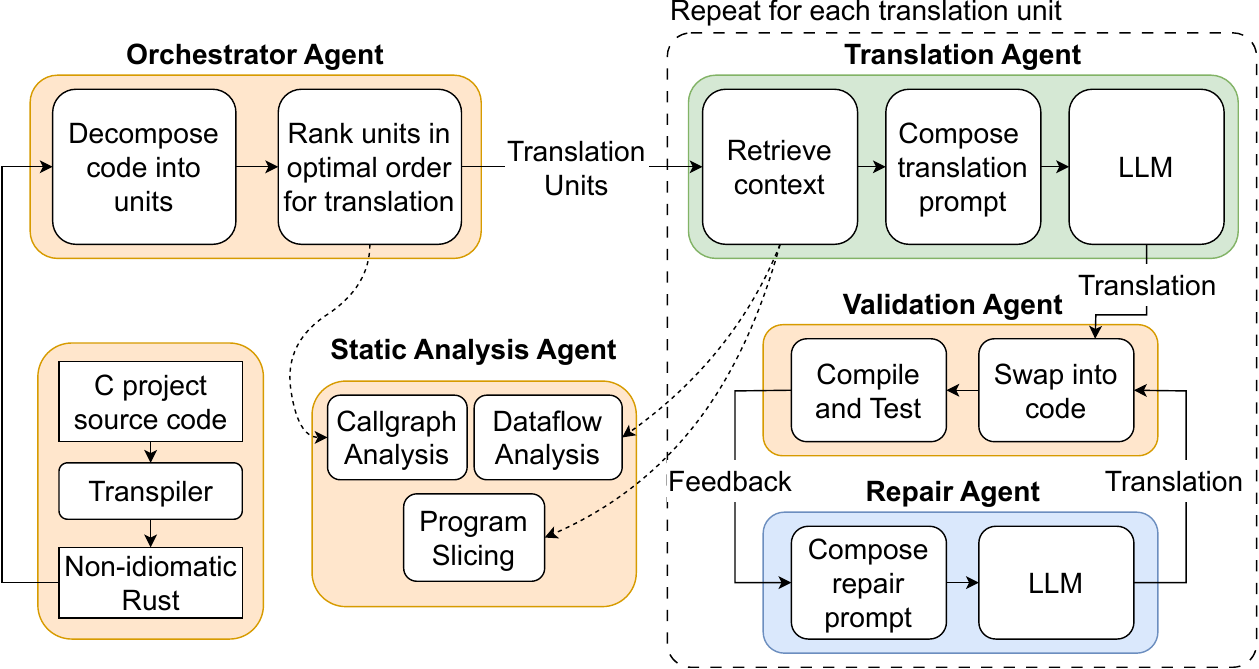}
    \caption{An overview of our system, \tool, to translate a C program to safer Rust. We first run the C2Rust transpiler to get non-idiomatic Rust, then decompose the program into chunks. For each chunk, we slice the program to extract relevant context information and dataflow information, and compose this into a translation prompt for an LLM. The LLM produces safer Rust, which we swap back in and test end-to-end. If the program doesn't compile or pass test cases, we give the LLM feedback and repeat.}
    \label{fig:sysdiagram}
\end{figure*}

\noindent
\textbf{Summary of Results:}  We evaluate on two datasets - one contributed by us, and one from a prior work, Laertes \cite{emre2021translating}.

\bi
\item On our dataset, \tool is able to reduce the amount of unsafe code by up to 28\%. We are also able to substantially reduce the occurrence of \textit{raw pointers}, a datatype that is exclusively associated with unsafe Rust.
\item On the Laertes dataset, \tool outperforms two selected baseline approaches on a majority of the programs for all of our metrics.
\item We show through an ablation study that each of the components of our design contributes to its overall performance.
\ei

\noindent
\textbf{Our Contributions:} In summary, our contributions in this paper as as follows:
\be
\item We adopt a novel approach of using C2Rust transpiled code as a bridge between C code and LLM-generated safe Rust, combining the strengths of rule-based transpiler approaches and LLMs. We traverse the program parse tree to decompose a large program into smaller units that can be individually translated by an LLM. We then augment each unit with context information derived from static program analysis. We test and validate the LLM-generated translations, and use feedback-guided repair to iteratively refine them.
\item We contribute a benchmark dataset of 7 programs consisting of utilities from the GNU Coreutils project. Each program comes with end-to-end system-level test cases, enabling us to evaluate correctness.
\item The ecosystem of static analysis tools for Rust is rather limited, and so we had to implement most of our analyses as Rust compiler passes, directly using the Rust compiler's intermediate representations. This lays the groundwork to build future Rust static analysis tools. For instance, we build program slicing and callgraph analysis algorithms, which are quintessential static analysis techniques. We make all of this code publicly available on GitHub\footnote{\url{https://github.com/vikramnitin9/c2saferrust}}.
\ee

\section{Background}

\subsection{Safety and Unsafety in Rust}

Rust is a systems programming language designed to offer memory safety guarantees while providing performance comparable to C and C++. In contrast, C and C++ are widely used but prone to memory errors and security vulnerabilities~\cite{durumeric2014matter, leveson1993investigation, seacord2013secure}. Rust achieves safety through a strict ownership model: each value in memory has a single owner, and deallocation occurs automatically when the owner goes out of scope. Values may be \textit{borrowed} by other variables, but such borrows are annotated with lifetimes, which the compiler tracks to ensure validity. A common pattern is for functions to borrow values from callers for the duration of the function and return ownership afterward.

Rust’s type system distinguishes between \textit{immutable} (\TT{\&T}) and \textit{mutable} (\TT{\&mut T}) borrows, enforcing exclusivity rules to prevent data races. Multiple immutable borrows may coexist, as they do not modify state, but a mutable borrow requires exclusive access, disallowing any simultaneous borrows. The borrow checker ensures statically that these constraints hold, resulting in memory safety without runtime overhead~\cite{weiss2019oxide}.

While these guarantees eliminate entire classes of bugs, they also make certain programming tasks more difficult. In particular, designing data structures that rely on shared ownership or cyclic references (such as graphs) becomes non-trivial, and careful design is needed to avoid lifetime conflicts. Safety enforcement may also introduce bounds checks or borrow checking overhead and complicate interactions with low-level system components or foreign libraries~\cite{astrauskas2020programmers}. To support such cases, Rust exposes the \TT{unsafe} keyword, permitting raw pointer manipulation, mutable static memory, and calls to unsafe functions (including FFI and compiler intrinsics)~\cite{rustreference}. Unsafe enables performance-critical or low-level operations but transfers responsibility for correctness from the compiler to the programmer.

The option to selectively use \TT{unsafe} while retaining strong guarantees elsewhere has driven efforts to port legacy C and C++ systems into Rust, aiming to reduce vulnerability surfaces associated with manual memory management~\cite{seacord2013secure}. However, this process is often labor-intensive. The permissive memory model in C—with raw pointers, unchecked array access, and no concept of ownership—makes straightforward translation into Rust’s stricter semantics challenging and error-prone~\cite{li2024translating}.

\subsection{Translating C to Rust using Transpilation}

A common approach to porting C code is the use of source-to-source compilers, or \textit{transpilers}, which aim to transform code from one language into another while preserving semantics. Tools such as C2Rust~\cite{c2rust} translate C constructs into Rust systematically and can automatically produce compilable Rust code.

\noindent\textbf{Challenges.}~
Despite providing a starting point, transpilation often produces code that compiles but does not fully benefit from Rust’s safety guarantees. Transpilers map raw pointers to \TT{*const} and \TT{*mut}, frequently requiring unsafe dereferencing, and global variables in C typically become \TT{static mut}, which also demands \TT{unsafe} access. Common C idioms such as pointer arithmetic, manual memory management, or type reinterpretation via \TT{transmute} reintroduce the possibility of undefined behavior. Moreover, transpiled code often retains low-level \TT{malloc}/\TT{free} semantics rather than adopting idiomatic Rust abstractions like \TT{Box}, \TT{Rc}, or \TT{Arc}, increasing reliance on unsafe and complicating verification of memory safety.

\subsection{Programming language translation with Large Language Models (LLMs)}

Large Language Models offer an emerging alternative for code translation. Their ability to generate code with contextual reasoning allows them to restructure programs more flexibly than rule-based transpilers, and they may incorporate safe coding idioms learned from training corpora~\cite{pan2024lost, yang2024exploring, ibrahimzada2024repository}. LLMs have shown promising results on structured benchmarks such as CodeNet~\cite{puri2021codenet}, Avatar~\cite{ahmad2021avatar}, and EvalPlus~\cite{evalplus}, where problems are self-contained and relatively small.

However, studies demonstrate that translating real-world codebases remains difficult. LLMs often struggle to maintain semantic equivalence across multi-file projects, preserve API interactions, or reason about complex type and memory relationships~\cite{ibrahimzada2024repository}. These limitations manifest as type mismatches, missing edge cases, broken dependencies, or unsafe substitutions. As a result, improving robustness and safety in LLM-based translation remains an active research direction~\cite{pan2024lost}.
\section{Methodology}

\subsection{Overview}
\tool is a neuro-symbolic framework for translating C programs to their safer Rust counterparts, focusing on preserving the intent of the original code. 
\tool is designed to operate on complex, real-world C projects, often comprising thousands of lines of code and tens of interdependent files. The translation process prioritizes functional correctness, ensuring that the resulting Rust code behaves as expected under the test cases provided by the user.

We leverage on-demand static analysis and LLMs for translation, and differential testing for validating the translated code. However, given the size of typical C project, directly translating an entire C project to Rust in one step is infeasible. We need to come-up with a strategy to translate and validate smaller pieces of code in a step-by-step manner. Thus, we need a framework that allows partial C and partial (translated) Rust code to coexist, as we need to validate the accuracy of the translation by running this hybrid code. To this end, we adopted the C2Rust framework, which transpiles a given C code repository into an equivalent unsafe Rust repository. While the C2Rust-transpiled code retains most of C's `unsafe` properties and original intent, it effectively acts as a Rust wrapper, enabling the unsafe code to execute within a Rust environment. Our translation process begins from this starting point, progressively transforming the unsafe Rust into safer Rust.

 This unsafe Rust code is then divided into smaller units, called {\em translation units}, using program graphs extracted through symbolic reasoning. Each unit is translated into safer Rust using an LLM, which generates safer code while symbolic reasoning ensures accuracy. Compilation and test cases identify errors, prompting targeted analysis to refine context and guide effective repairs. This iterative approach balances efficiency and accuracy, progressively enhancing translation quality to suit the structure and requirements of each project. An overview of the system is shown in \Cref{fig:sysdiagram}. \tool consists of the following components:

\be
\item \textbf{Translation Orchestrator (\Cref{sec:orchestration}):} This component, aided by static analysis, orchestrates translation. It breaks the project into smaller translation units that fit within the LLM’s context. It then selects translation units in the optimal order for translation. Translation units are not independent - when one unit is translated, the side effects have to be propagated to all other affected units. The optimal order for translation is \textit{moving up} the chain of propagated side effects, \ie translating a unit \textit{before} translating the units to which its side effects are propagated. This ensures that while the LLM is translating a unit, it can also assimilate the side effects propagated from other units. We will discuss more about this with concrete examples in \Cref{sec:orchestration}.
\item \textbf{Static Analyzer (\Cref{sec:slicing}):} We collect key dependency data (e.g., call graph, dataflow analysis) using traditional program analysis techniques. Since our translation process is incremental, the program changes at every translation step, and the static analyzer must be re-run each time.
\item \textbf{Translation (\Cref{sec:translation}):} The translation prompt is composed of the code component to be translated, augmented with information derived from static analysis. This information describes the surrounding context of the code segment and its dependencies, which guides the LLM in translating the unsafe Rust code to safe Rust.
\item \textbf{Validator (\Cref{sec:translation}):} This component acts as a regression testing framework, validating the translation of each unit by first compiling the new code, and then running system-level tests provided with the original C code. The safe Rust components replace their unsafe counterparts progressively, allowing safe and unsafe Rust to coexist during validation.

\item \textbf{Repair (\Cref{sec:translation}):} After a unit is translated, if there an error in validation, then the repair component of our system collects feedback from compilation and execution. Then, it prompts the LLM with this feedback and asks it to re-generate the translation of the unit.
\ee

For each translation unit, the cycle of Validation and Repair continues until the validation succeeds, or we exceed $N$ attempts (where $N$ is a parameter of our system). If we are unable to translate a unit within $N$ attempts, we leave it as its original unsafe Rust version, thus maintaining overall correctness.
In this manner, we go through the entire unsafe Rust program, processing each translation unit in order. Once we translate (or attempt to translate) a unit, we do not come back to it later. In the remainder of this section, we describe each of the components of our system in detail.


\subsection{Translation Orchestrator}
\label{sec:orchestration}

\subsubsection{Decomposing the program into translation units}
\label{sec:decomposition}
Our goal is to translate the entire program into safe Rust; however, we are limited by the size of the LLM context window. Further, even if the input and output fit within the context window, LLMs are more prone to hallucinations and translation errors when the input is long. Our solution is to decompose the program into translation units of less than $L$ lines each, and translate these units individually.

If a function has fewer than $L$ lines, we extract the whole function as a translation unit. Otherwise, we divide the function into multiple translation units of fewer than $L$ lines, which are possibly nested. At a high level, we traverse the AST of the function in a DFS post-order traversal, and visit only the sub-trees that are greater than $L$ lines long. In each such sub-tree, we form groups of sibling nodes and extract the corresponding statements in each group as a translation unit. Once a unit is extracted, it is removed from the AST. In this way, as we move up the AST, we ensure that every subtree has fewer than $L$ lines. By induction, at the end of the traversal, the root tree will also have fewer than $L$ lines. For the full algorithm, please refer to the appendix. 

\begin{figure*}
    \centering
    \includegraphics[width=0.75\linewidth]{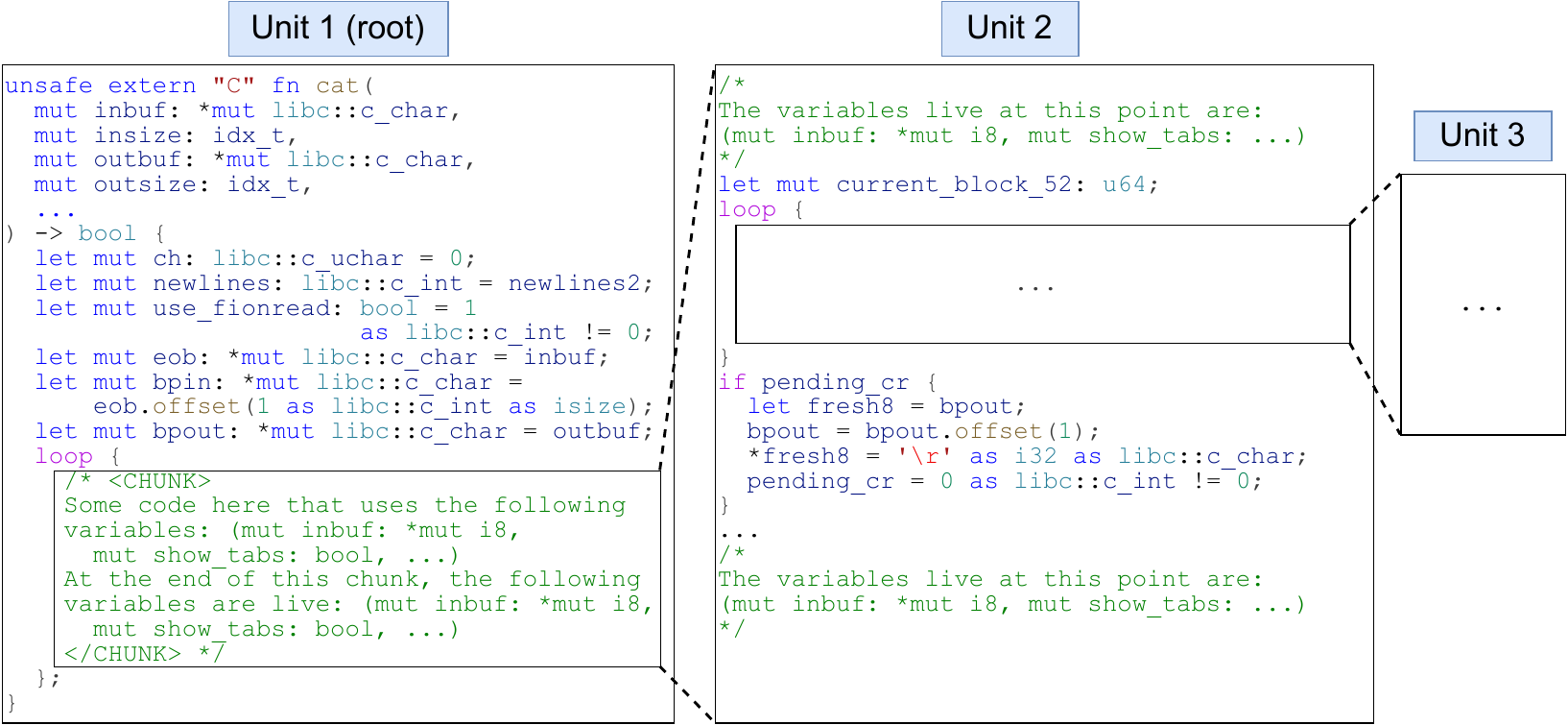}
    \caption{An example of the output of our decomposition algorithm as described in \Cref{sec:decomposition}, applied to the \TT{cat} function from the \TT{cat} program. Translation Unit 1 is the ``root'' unit that represents the function as a whole, and all other units are nested within it. The units are annotated with comments showing live-in and live-out information, which is described in \Cref{sec:slicing}.}
    \label{fig:decomposition}
\end{figure*}

Each function will be decomposed into $n$ translation units. These units are of two kinds:
\bi
\item \textbf{Root translation unit:} One out of the $n$ translation units corresponds to the entire function, and contains all the other units within it. It also includes the function signature.
\item \textbf{Inner translation units:} The remaining $n-1$ translation units are nested within each other or within the root units.
\ei
An example of this is shown in \Cref{fig:decomposition}. Note that this decomposition does not include code that is outside of functions, like structure and enum definitions. Translating these into safe Rust is a non-trivial problem, and is out of the scope of our paper.

\subsubsection{Arranging translation units in an optimal order}
\label{sec:ordering}

The order in which we convert unsafe Rust translation units to safe Rust makes a difference to the quality of translation. Let us say we are converting a function \TT{A} that calls an unsafe function \TT{B}. Then even if we ``clean up'' the body of \TT{A} to use safe Rust for the most part, \textit{it must still rely on \TT{unsafe} blocks to call \TT{B}}. Further, if \TT{B} has arguments or return values that are raw pointers, then \TT{A} will have to use \TT{unsafe} code to manipulate these values. However, once \TT{B} is ``cleaned up'' to use only safe and safer Rust types, then we can convert \TT{A} to safe Rust in its entirety. So each function should ideally be processed \textit{before its callers}.

\begin{figure}
    \centering
    \includegraphics[width=\linewidth]{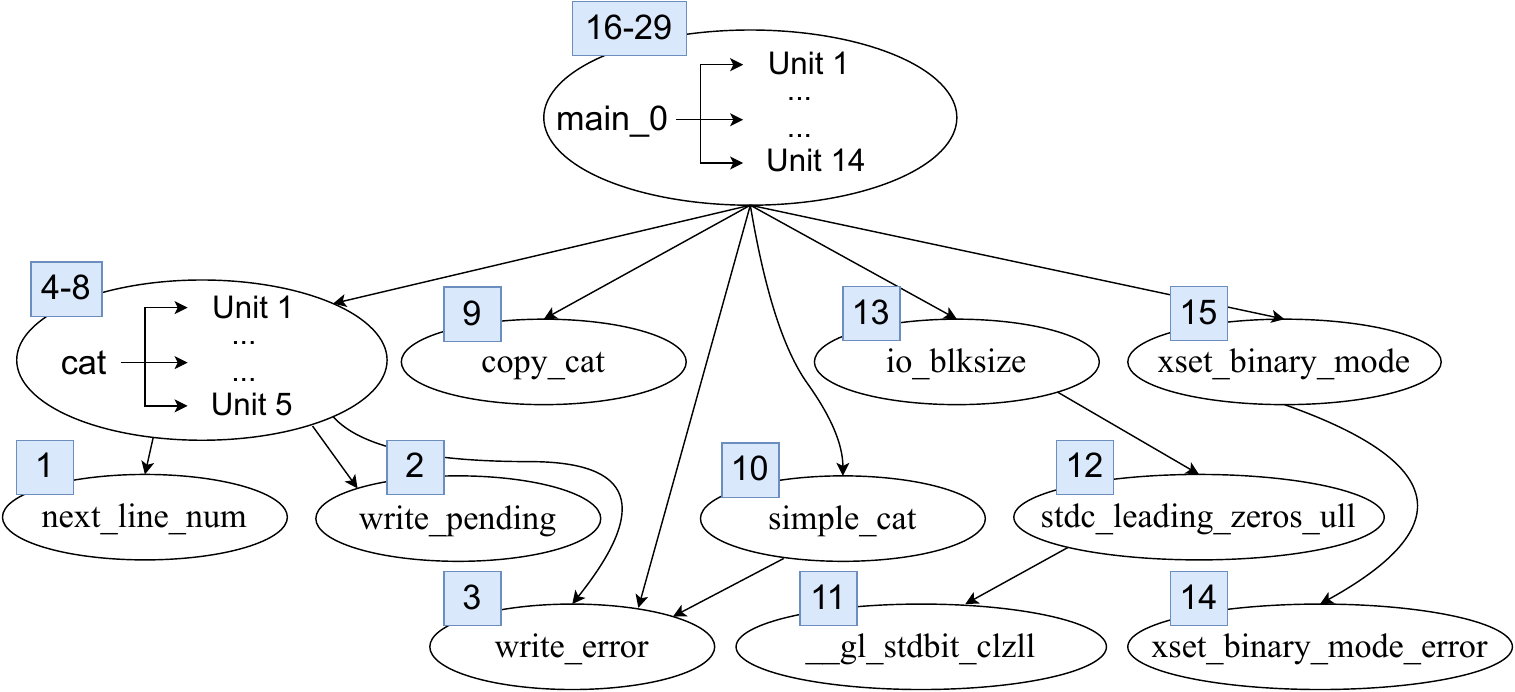}
    \caption{A portion of the callgraph of \TT{cat}. Each node denotes a function, and the edges denote calls. The functions are sorted in a reverse topological order. The numbers next to each function denote their positional index in the ordering of translation units. \TT{main\_0} and \TT{cat} are associated with a \textit{range} of indices (4-8 and 16-29) because they have multiple translation units within them.}
    \label{fig:callgraph}
\end{figure}

To ensure this, we build a call-graph of the program and compute its \textit{weakly connected components}. A weakly connected component (WCC) is a subgraph such that, if all its directed edges are converted to undirected edges, all its nodes have paths between each other. Functions within each WCC need to be ordered such that callee comes before caller. So we do a \textit{reverse topological sort} of each WCC. If a WCC has a cycle, there is no topological ordering. In this case, we start at an arbitrary node in this WCC and order the nodes according to a depth-first post-order traversal.

This gives us an ordering of functions \textit{within} each WCC of the call graph. The order of functions \textit{between} two WCCs doesn't matter, so we process the WCCs one at a time in an arbitrary order to create the full ordering of all functions in the program. Further, each function could contain multiple translation units. As we will discuss in \Cref{sec:translation}, the order of translation units \textit{within} a function does not matter for our current translation strategy. However, we ensure that all the translation units corresponding to a single function are contiguous in the translation order.

\subsection{Static Analyzer}
\label{sec:slicing}

We need to annotate each translation unit with enough context to enable the LLM to translate the code well. Further, each unit cannot be translated in isolation, because it could have ripple effects on other regions of the code. Our solution is to translate all these affected regions \textit{simultaneously}.

We obtain this information through static analysis of the program. The information we derive differs for a root translation unit and an inner translation unit.

A \textit{root} translation unit will contain the function header and signature, along with part of the function body. We cannot translate this in isolation, since its parameters or return type might change in this process (\eg raw pointers might become borrows). These changes are actually desirable, because we need to modify function signatures to transition away from unsafe Rust idioms like raw pointers.

When a function signature changes, this change will spread to all of its call sites. So we ask the LLM to translate a function \textit{together with all of its call sites}. We also add other contextual information to help the LLM, such as global variables and \TT{use} statements. Thus, each slice for a root translation unit consists of:
\be
\item[\circled{1}] The source code of the translation unit
\item[\circled{2}] The function's call sites
\item[\circled{3}] Global variables used in the function
\item[\circled{4}] Module imports used in the file
\ee

On the other hand, for an \textit{inner} unit, we need to tell the LLM about the variables that are used and created within the unit. We use dataflow analysis to get the \textbf{live-in} and \textbf{live-out} variables for the unit. So each slice consists of:
\be
\item[\circled{1}] The source code of the translation unit
\item[\circled{2}] Its live-in and live-out variables
\item[\circled{3}] Global variables used in the translation unit
\item[\circled{4}] Module imports used in the file
\ee

\begin{figure}
    \centering
    \includegraphics[width=\linewidth]{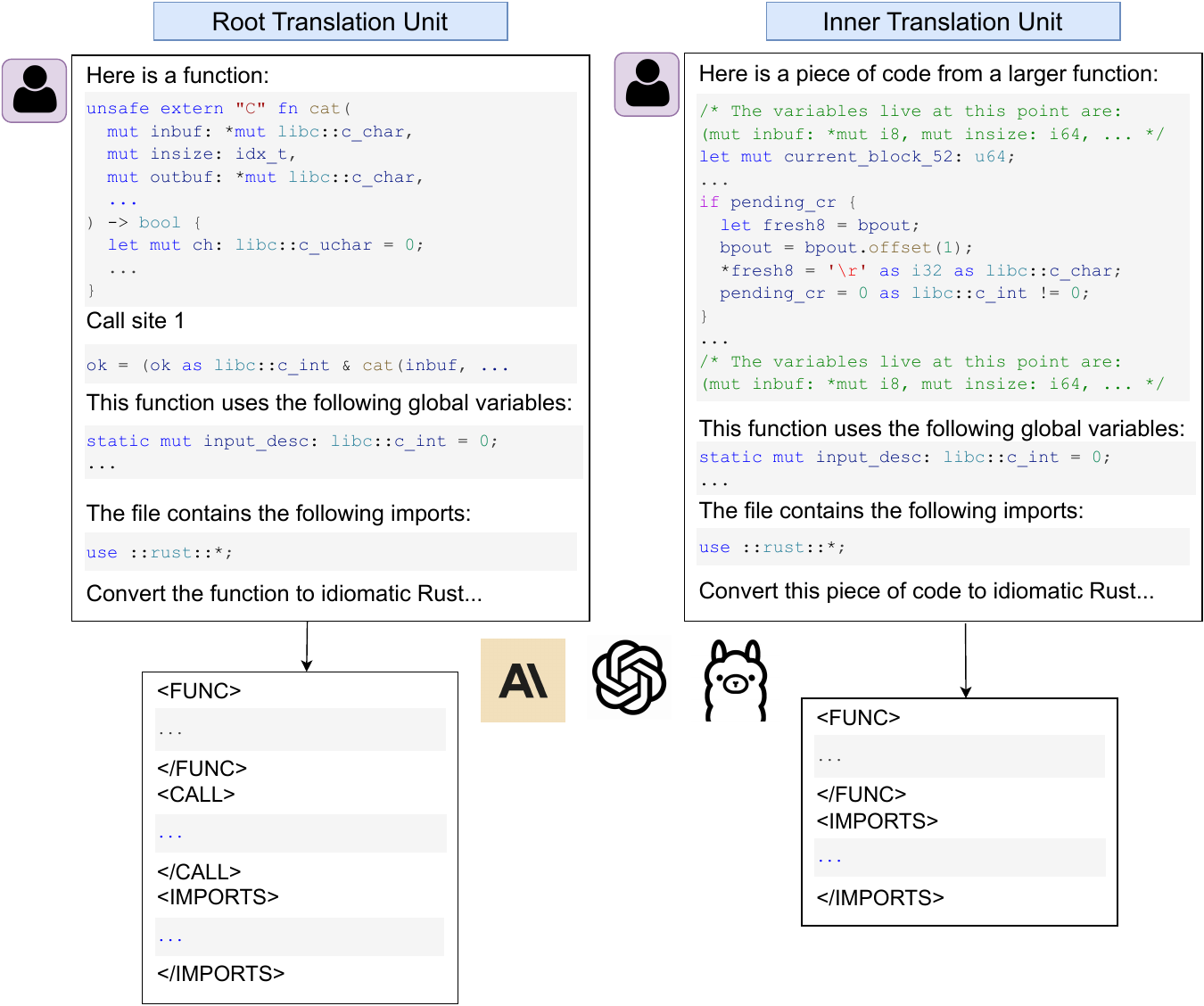}
    \caption{Translating units with an LLM. For the root translation unit (left), we generate the translation of the function body plus all of its call sites simultaneously. For the inner translation unit (right), we translate just the code contained within it. In both cases, we also allow the model to specify any imports it needs. \rebuttal{[Reviewer 3]}{Full prompts are in the Appendix.}}
    \label{fig:prompt}
\end{figure}

\Cref{fig:prompt} shows slices for a root translation unit and an inner translation unit, composed into an LLM prompt.

\subsection{Translation, Validation and Repair}
\label{sec:translation}

We now attempt to translate each translation unit from unsafe Rust to safe, idiomatic Rust, using an LLM. We pass each unit to an LLM, along with its context slice as described in \Cref{sec:slicing}. We instruct the LLM to convert the translation unit to idiomatic Rust. If the unit is a root translation unit of a function and its context contains call sites, we instruct the LLM to translate the call sites too. We specify a particular format for the LLM response that allows us to easily parse it and separate it into function and call-site translations. If the unit is an inner translation unit, we ask the LLM to ensure that the live-in and live-out variables at the beginning and end of the unit are not affected. This means that translating such a translation unit has no side-effects on any other part of the program. This substantiates our claim in \Cref{sec:ordering} that the order in which we translate the units within a function does not matter.

Sometimes, the translated unit needs additional \TT{use} statements to work; we allow the LLM to specify this too in its response. An example of a prompt and LLM response is shown in \Cref{fig:prompt}.

We then \textbf{swap out} the original spans of code and \textbf{swap in} the LLM-generated code. We attempt to compile the code and run the test suite. If we encounter an error, we provide it as feedback to the LLM in the same conversation, and ask it to re-generate the entire code, including the function body, call sites and imports. We repeat this for up to $N$ attempts (where $N$ is a parameter of our technique). If one of these $N$ attempts succeeds, we move on to the next function; if all $N$ attempts fail, we restore the original version of the function.

In this manner, we go through the entire program, processing each translation unit exactly once. Throughout this process, we maintain correctness modulo the test cases. Our final program will still have some unsafe functions, but these can be converted to safe Rust by a human expert.
\section{Experimental Setup}

\subsection{Datasets}

We contribute a benchmark dataset of 7 C programs collected from GNU coreutils\footnote{\url{https://github.com/coreutils/coreutils}}. These are the basic file, shell and text manipulation utilities of the GNU OS, and include common shell commands like \TT{ls} and \TT{cat}. We chose coreutils because it is a large, mature C code project, with many constituent programs of varying sizes. Further, there is an ongoing community effort to manually rewrite coreutils in Rust\footnote{\url{https://github.com/uutils/coreutils}}, which means that a) translating coreutils to Rust is an important real-world problem, and b) future researchers using our benchmark can compare the quality of their automatically-generated idiomatic Rust code with human-written Rust code. Finally, each benchmark in coreutils comes with end-to-end system-level test cases, which have been constructed using fuzzing to ensure high coverage\footnote{\url{https://www.pixelbeat.org/docs/
coreutils-testing.html}}.

The 7 programs we picked from coreutils are \TT{split}, \TT{pwd}, \TT{cat}, \TT{truncate}, \TT{uniq}, \TT{tail}, and \TT{head}. Running C2Rust on these programs was not straightforward, because C2Rust does not work directly with Makefiles of complex programs. Coreutils uses a compilation pattern where some object files are bundled into static libraries, and those are then linked to the necessary executables. This compilation pattern makes it difficult for C2Rust to create a compilation database, which is necessary for further analysis.

To make this problem tractable, we create simpler individual Makefiles for each of the 7 programs, including only the \TT{.c} and \TT{.h} files that are strictly necessary for each program. We collect the source files for each program iteratively. Starting with the main C file for that program (\eg~ \TT{split.c} for \TT{split}), we collect the necessary header files and attempt to build the program. If the linking step fails due to missing symbols, we collect the necessary source files that define these symbols and attempt to build the program again. We repeat until the executable is built successfully. The simpler Makefile is then fed to C2Rust.

The statistics of these 7 programs are shown in \Cref{tab:dataset}. In addition to these 7 programs from coreutils, we also use benchmarks from a prior work, Laertes \cite{emre2021translating}. \rebuttal{[Reviewer 3 - why only 10?]}{Since evaluating on each additional benchmark incurs a time cost, we selected a subset of 10 out of their 17 benchmarks, which we believe is sufficient for the purposes of a fair evaluation.} These benchmarks do not come with end-to-end test cases. Although some of the utilities have \textit{unit} tests, this is not useful because we cannot run C unit tests on Rust programs.

\begin{table}[ht]
    \centering
    \resizebox{\linewidth}{!} {
    \begin{tabular}{l|r|c|c|c|c|c}
    \toprule
         & \textbf{Program} & \textbf{\# LoC} & \textbf{\# Fn} & \makecell{\rebuttal{}{\textbf{\# Cov.}}\\ \rebuttal{}{\textbf{Fn}}} & \makecell{\textbf{Max}\\\textbf{Fn Len}} & \makecell{\textbf{\textbf{Mean}}\\\textbf{Fn Len}} \\
         \bottomrule
\multirow{7}{*}{\rot{90}{\textbf{\textsc{Coreutils}}}} & split & 13848 & 207 & \rebuttal{}{73} & 1634 & 52.95 \\
& pwd & 5859 & 127 & \rebuttal{}{16} & 902 & 33.02 \\
& cat & 7460 & 166 & \rebuttal{}{37} & 1006 & 33.84 \\
& truncate & 7181 & 124 & \rebuttal{}{33} & 902 & 44.65 \\
& uniq & 8299 & 167 & \rebuttal{}{34} & 902 & 36.26 \\
& tail & 14423 & 266 & \rebuttal{}{76} & 902 & 43.81 \\
& head & 8047 & 153 & \rebuttal{}{45} & 902 & 40.77 \\
         \midrule
\multirow{10}{*}{\rot{90}{\textbf{\textsc{Laertes}}}} & tulipindicators & 44486 & 234 & -- & 306 & 32.10 \\
& xzoom & 2524 & 11 & -- & 536 & 134.09 \\
& genann & 2084 & 32 & -- & 201 & 52.44 \\
& optipng & 95560 & 576 & -- & 42124 & 143.63 \\
& urlparser & 1118 & 22 & -- & 409 & 38.41 \\
& lil & 5400 & 160 & -- & 283 & 30.44 \\
& qsort & 41 & 3 & -- & 17 & 10.67 \\
& snudown & 6521 & 92 & -- & 792 & 63.53 \\
& grabc & 1046 & 7 & -- & 61 & 31.57 \\
& bzip2 & 43374 & 128 & -- & 3289 & 98.84 \\
        \bottomrule
    \end{tabular}
    }
    \caption{The statistics of the 7 programs in our coreutils benchmark dataset, plus the 10 programs from Laertes. \revision{``\# Cov. Fn'' is the number of functions covered by tests.}}
    \label{tab:dataset}
\end{table}

\subsection{Baselines}

We evaluate against two previous approaches for making C2Rust output more natural.
\bi
\item \textbf{Laertes \cite{emre2021translating}:} Laertes uses a set of simple transformation rules to rewrite each raw pointer as a borrow, and then uses compiler feedback to fix errors until the program compiles.

\item \textbf{CROWN \cite{zhang2023ownership}:} CROWN derives ownership constraints for each raw pointer based on their usage patterns, and uses an SMT solver to solve these constraints and transform each raw pointer into a borrow.
\ei

\subsection{Implementation}
We implement \tool on the \TT{nightly-2022-08-08} version of Rust. It is implemented as a \TT{cargo} subcommand, so it can be run on full Cargo packages. We use the Rust compiler's High-level Intermediate Representation (HIR) to construct the program call graph and to decompose the program into chunks. The AST traversal in \Cref{alg:decomposition} is implemented as an instance of the \TT{rustc\_hir::intravisit::Visitor} trait. The slicing described in \Cref{sec:slicing} is also performed on the HIR.

We use the Mid-level Intermediate Representation (MIR) to solve the dataflow analysis for each chunk as described in \Cref{sec:slicing}.
We use the MIR debug information to find the first and last basic blocks in a chunk, and get the \textit{live-in} variables of the \textit{first} block and the \textit{live-out} variables of the \textit{last} block. These become the live-in and live-out variables of the chunk.

For all our experiments, the LLM used is \TT{gpt4o-mini} by OpenAI. We fixed the maximum chunk size, $L$, at 150 lines, and the number of LLM feedback iterations, $N$, at 5. The smallest baseline, \TT{qsort}, took under 5 mins to translate, and the largest, \TT{optipng}, took nearly 14 hours. \rebuttal{[Reviewer 3 - cost]}{More details about LLM calls and costs are in the Appendix.}

\subsection{Evaluation}
\label{sect:eval}
We use two metrics from prior work to evaluate the quality of translation:
\bi
\item \textbf{Raw Pointer Declarations:} The number of raw pointer declarations, \TT{*const T} or \TT{*mut T}, in the entire program. This includes declarations in function signatures and within structures. This metric was used to evaluate Laertes \cite{emre2021translating}.

\item \textbf{Raw Pointer Dereferences:} The number of raw pointer dereferences throughout the program. This includes all dereferences of the form \TT{*<expr>} where \TT{<expr>} evaluates to a raw pointer type (verified by type checking). This metric was also used to evaluate Laertes \cite{emre2021translating}.
\ei 

Raw pointers are not a feature of idiomatic Rust, and the fewer raw pointers in a program, the more natural it is. Further, dereferencing raw pointers is an unsafe operation which could lead to segmentation faults and other undefined behavior if not used correctly. So for both of the above metrics, a lower value indicates safer code. We further analyze unsafe blocks by investigating their length and types of statements, through the following three metrics:
\bi
\item \textbf{Unsafe Lines of Code:} The total lines of code (LOC) contained within all \TT{unsafe} blocks in the program. This includes the bodies of unsafe functions, which are also treated as unsafe blocks for this metric. A lower value for unsafe LOC indicates a more natural and idiomatic Rust program, as it minimizes the reliance on unsafe constructs.
\item \textbf{Unsafe Type Casts:} Type casts, within unsafe blocks, especially those involving raw pointers (such as \TT{ptr as *const u8}) are a common pattern which bypass Rust's strict type system, allowing for potential misuse and may result in undefined behavior if the cast is used incorrectly. 
\item \textbf{Unsafe Call Expressions:} Function calls within unsafe blocks, such as \TT{std::alloc::alloc(layout)} or other custom unsafe functions, allow for a direct interaction with low-level operations. This can be dangerous when encapsulated in unsafe blocks because they bypass Rust's safety guarantees and potentially lead to memory corruption, dangling pointers, or other undefined behavior.
\ei 
\section{Results}
\subsection{Effectiveness of \tool in reducing unsafe constructs (RQ1)}
\label{sec:coreutils}
\noindent\textit{\textbf{Motivation:}}
In the first research question, we try to evaluate how effectively \tool can transform C2Rust's unsafe code into safe and idiomatic Rust.

\noindent\textit{\textbf{Evaluation:}} We run \tool on each program in our benchmark dataset, which was compiled from \texttt{coreutils}. For each program, we calculated the following metrics both before and after translation (see \cref{sect:eval} for detailed information): (1) the count of raw pointer declarations, (2) the number of raw pointer dereferences, and (3) the extent of unsafe lines of code. Additionally, our analysis was expanded to encompass measurements of (1) unsafe type casts and (2) unsafe call expressions.
\begin{table*}[t!]
\caption{Translating CoreUtils to safer Rust using \tool. Overall results suggest \tool offers enhanced safety by reducing reliance on raw pointers and unsafe operations.}
\label{tab:coreutils}
\begin{minipage}{\linewidth}
\renewcommand{\baselinestretch}{1.25}\selectfont
\centering
\resizebox{0.67\linewidth}{!}{%
\begin{tabular}{lrrrrrrrrrrrrrrr}
\clineB{2-16}{2}
\multicolumn{1}{lV{2}}{\multirow{3}{*}{}} &
  \multicolumn{3}{cV{2}}{Raw Pointer} &
  \multicolumn{3}{cV{2}}{Raw Pointer} &
  \multicolumn{3}{cV{2}}{Unsafe} &
  \multicolumn{3}{cV{2}}{Unsafe Call}\bigstrut[t] &
  \multicolumn{3}{cV{2}}{Unsafe}
  
  \\
\multicolumn{1}{lV{2}}{} &
  \multicolumn{3}{cV{2}}{Declarations ($\nabla$)} &
  \multicolumn{3}{cV{2}}{Deferences  ($\nabla$)} &
  \multicolumn{3}{cV{2}}{Lines ($\nabla$)} & 
  \multicolumn{3}{cV{2}}{Expressions  ($\nabla$)} &
  \multicolumn{3}{cV{2}}{Type Casts ($\nabla$)}
  \bigstrut[b]\\
  \clineB{2-16}{2}
\multicolumn{1}{lV{2}}{} &
  \multicolumn{1}{rV{2}}{\rotatebox{90}{Before~~}} &
  \multicolumn{1}{rV{2}}{{\rotatebox{90}{After}}} &
  \multicolumn{1}{cV{2}}{$\Delta\%$} &
  \multicolumn{1}{rV{2}}{\rotatebox{90}{Before~~}} &
  \multicolumn{1}{rV{2}}{{\rotatebox{90}{After}}} &
  \multicolumn{1}{cV{2}}{$\Delta\%$} &
  \multicolumn{1}{rV{2}}{\rotatebox{90}{Before~~}} &
  \multicolumn{1}{rV{2}}{{\rotatebox{90}{After}}} &
  \multicolumn{1}{cV{2}}{$\Delta\%$} &
  \multicolumn{1}{rV{2}}{\rotatebox{90}{Before~~}} &
  \multicolumn{1}{rV{2}}{{\rotatebox{90}{After}}} &
  \multicolumn{1}{cV{2}}{$\Delta\%$} &
  \multicolumn{1}{rV{2}}{\rotatebox{90}{Before~~}} &
  \multicolumn{1}{rV{2}}{{\rotatebox{90}{After}}} &
  \multicolumn{1}{cV{2}}{$\Delta\%$}
  \bigstrut\\
  \clineB{2-16}{2}
 &
  \multicolumn{1}{l}{} &
  \multicolumn{1}{l}{} &
  \multicolumn{1}{l}{} &
  \multicolumn{1}{l}{} &
  \multicolumn{1}{l}{} &
  \multicolumn{1}{l}{} &
  \multicolumn{1}{l}{} &
  \multicolumn{1}{l}{} &
  \multicolumn{1}{l}{} &
  \multicolumn{1}{l}{} &
  \multicolumn{1}{l}{} &
  \multicolumn{1}{l}{} &
  \multicolumn{1}{l}{} &
  \multicolumn{1}{l}{}
  \bigstrut\\[-1.33em]
  \hlineB{2}
\multicolumn{1}{V{2}lV{2}}{split}
& \multicolumn{1}{rV{2}}{252}
& \multicolumn{1}{rV{2}}{214}
& \multicolumn{1}{rV{2}}{\cellcolor{blue!12}15}
& \multicolumn{1}{rV{2}}{656}
& \multicolumn{1}{rV{2}}{540}
& \multicolumn{1}{rV{2}}{\cellcolor{blue!12}18}
& \multicolumn{1}{rV{2}}{11324}
& \multicolumn{1}{rV{2}}{9250}
& \multicolumn{1}{rV{2}}{\cellcolor{blue!12}18}
& \multicolumn{1}{rV{2}}{2353}
& \multicolumn{1}{rV{2}}{2212}
& \multicolumn{1}{rV{2}}{\cellcolor{blue!12}6}
& \multicolumn{1}{rV{2}}{5979}
& \multicolumn{1}{rV{2}}{4663}
& \multicolumn{1}{rV{2}}{\cellcolor{blue!12}22}
\bigstrut\\
\multicolumn{1}{V{2}lV{2}}{pwd}
& \multicolumn{1}{rV{2}}{164}
& \multicolumn{1}{rV{2}}{129}
& \multicolumn{1}{rV{2}}{\cellcolor{blue!12}21}
& \multicolumn{1}{rV{2}}{295}
& \multicolumn{1}{rV{2}}{225}
& \multicolumn{1}{rV{2}}{\cellcolor{blue!12}24}
& \multicolumn{1}{rV{2}}{4201}
& \multicolumn{1}{rV{2}}{3151}
& \multicolumn{1}{rV{2}}{\cellcolor{blue!12}25}
& \multicolumn{1}{rV{2}}{875}
& \multicolumn{1}{rV{2}}{871}
& \multicolumn{1}{rV{2}}{\cellcolor{blue!12}0}
& \multicolumn{1}{rV{2}}{2248}
& \multicolumn{1}{rV{2}}{1563}
& \multicolumn{1}{rV{2}}{\cellcolor{blue!12}30}
\bigstrut\\
\multicolumn{1}{V{2}lV{2}}{cat}
& \multicolumn{1}{rV{2}}{192}
& \multicolumn{1}{rV{2}}{120}
& \multicolumn{1}{rV{2}}{\cellcolor{blue!30}\textbf{38}}
& \multicolumn{1}{rV{2}}{317}
& \multicolumn{1}{rV{2}}{240}
& \multicolumn{1}{rV{2}}{\cellcolor{blue!12}24}
& \multicolumn{1}{rV{2}}{5625}
& \multicolumn{1}{rV{2}}{4189}
& \multicolumn{1}{rV{2}}{\cellcolor{blue!12}26}
& \multicolumn{1}{rV{2}}{1038}
& \multicolumn{1}{rV{2}}{912}
& \multicolumn{1}{rV{2}}{\cellcolor{blue!12}12}
& \multicolumn{1}{rV{2}}{3116}
& \multicolumn{1}{rV{2}}{1992}
& \multicolumn{1}{rV{2}}{\cellcolor{blue!12}36}
\bigstrut\\
\multicolumn{1}{V{2}lV{2}}{truncate}
& \multicolumn{1}{rV{2}}{156}
& \multicolumn{1}{rV{2}}{124}
& \multicolumn{1}{rV{2}}{\cellcolor{blue!12}21}
& \multicolumn{1}{rV{2}}{326}
& \multicolumn{1}{rV{2}}{263}
& \multicolumn{1}{rV{2}}{\cellcolor{blue!12}19}
& \multicolumn{1}{rV{2}}{5544}
& \multicolumn{1}{rV{2}}{4443}
& \multicolumn{1}{rV{2}}{\cellcolor{blue!12}20}
& \multicolumn{1}{rV{2}}{1040}
& \multicolumn{1}{rV{2}}{952}
& \multicolumn{1}{rV{2}}{\cellcolor{blue!12}8}
& \multicolumn{1}{rV{2}}{3357}
& \multicolumn{1}{rV{2}}{2521}
& \multicolumn{1}{rV{2}}{\cellcolor{blue!12}25}
\bigstrut\\
\multicolumn{1}{V{2}lV{2}}{uniq}
& \multicolumn{1}{rV{2}}{227}
& \multicolumn{1}{rV{2}}{166}
& \multicolumn{1}{rV{2}}{\cellcolor{blue!12}27}
& \multicolumn{1}{rV{2}}{343}
& \multicolumn{1}{rV{2}}{250}
& \multicolumn{1}{rV{2}}{\cellcolor{blue!30}\textbf{27}}
& \multicolumn{1}{rV{2}}{6066}
& \multicolumn{1}{rV{2}}{4397}
& \multicolumn{1}{rV{2}}{\cellcolor{blue!30}\textbf{28}}
& \multicolumn{1}{rV{2}}{1150}
& \multicolumn{1}{rV{2}}{1025}
& \multicolumn{1}{rV{2}}{\cellcolor{blue!12}11}
& \multicolumn{1}{rV{2}}{3590}
& \multicolumn{1}{rV{2}}{2340}
& \multicolumn{1}{rV{2}}{\cellcolor{blue!12}35}
\bigstrut\\
\multicolumn{1}{V{2}lV{2}}{tail}
& \multicolumn{1}{rV{2}}{389}
& \multicolumn{1}{rV{2}}{297}
& \multicolumn{1}{rV{2}}{\cellcolor{blue!12}24}
& \multicolumn{1}{rV{2}}{1092}
& \multicolumn{1}{rV{2}}{847}
& \multicolumn{1}{rV{2}}{\cellcolor{blue!12}22}
& \multicolumn{1}{rV{2}}{11663}
& \multicolumn{1}{rV{2}}{8818}
& \multicolumn{1}{rV{2}}{\cellcolor{blue!12}24}
& \multicolumn{1}{rV{2}}{2580}
& \multicolumn{1}{rV{2}}{2433}
& \multicolumn{1}{rV{2}}{\cellcolor{blue!12}6}
& \multicolumn{1}{rV{2}}{5869}
& \multicolumn{1}{rV{2}}{3739}
& \multicolumn{1}{rV{2}}{\cellcolor{blue!30}\textbf{36}}
\bigstrut\\
\multicolumn{1}{V{2}lV{2}}{head}
& \multicolumn{1}{rV{2}}{192}
& \multicolumn{1}{rV{2}}{141}
& \multicolumn{1}{rV{2}}{\cellcolor{blue!12}27}
& \multicolumn{1}{rV{2}}{442}
& \multicolumn{1}{rV{2}}{351}
& \multicolumn{1}{rV{2}}{\cellcolor{blue!12}21}
& \multicolumn{1}{rV{2}}{6245}
& \multicolumn{1}{rV{2}}{4699}
& \multicolumn{1}{rV{2}}{\cellcolor{blue!12}25}
& \multicolumn{1}{rV{2}}{1378}
& \multicolumn{1}{rV{2}}{1189}
& \multicolumn{1}{rV{2}}{\cellcolor{blue!30}\textbf{14}}
& \multicolumn{1}{rV{2}}{3488}
& \multicolumn{1}{rV{2}}{2464}
& \multicolumn{1}{rV{2}}{\cellcolor{blue!12}29}
\bigstrut\\
\hlineB{2}
 &
  \multicolumn{1}{l}{} &
  \multicolumn{1}{l}{} &
  \multicolumn{1}{l}{} &
  \multicolumn{1}{l}{} &
  \multicolumn{1}{l}{} &
  \multicolumn{1}{l}{} &
  \multicolumn{1}{l}{} &
  \multicolumn{1}{l}{} &
  \multicolumn{1}{l}{} &
  \multicolumn{1}{l}{} &
  \multicolumn{1}{l}{} &
  \multicolumn{1}{l}{} &
  \multicolumn{1}{l}{} &
  \multicolumn{1}{l}{}
  \bigstrut\\[-1.33em]
\hlineB{2}
\multicolumn{1}{V{2}lV{2}}{avg. $\Delta\%$}
& \multicolumn{1}{rV{2}}{}
& \multicolumn{1}{rV{2}}{}
& \multicolumn{1}{rV{2}}{\cellcolor{blue!12}25}
& \multicolumn{1}{rV{2}}{}
& \multicolumn{1}{rV{2}}{}
& \multicolumn{1}{rV{2}}{\cellcolor{blue!12}22}
& \multicolumn{1}{rV{2}}{}
& \multicolumn{1}{rV{2}}{}
& \multicolumn{1}{rV{2}}{\cellcolor{blue!12}24}
& \multicolumn{1}{rV{2}}{}
& \multicolumn{1}{rV{2}}{}
& \multicolumn{1}{rV{2}}{\cellcolor{blue!12}8}
& \multicolumn{1}{rV{2}}{}
& \multicolumn{1}{rV{2}}{}
& \multicolumn{1}{rV{2}}{\cellcolor{blue!12}30}
\bigstrut\\
\hlineB{2}
\end{tabular}%
}

\vspace{0.5em}
\subcaption{Our metrics for measuring safety of Rust code, computed across all applications before and after translation.}
\end{minipage}

\begin{minipage}{\linewidth}
\renewcommand{\baselinestretch}{1.25}\selectfont
\centering
\resizebox{0.67\linewidth}{!}{%
\begin{tabular}{lrrrrrrrrrrrrrrr}
\clineB{2-16}{2}
\multicolumn{1}{lV{2}}{\multirow{3}{*}{}} &
  \multicolumn{3}{cV{2}}{Raw Pointer} &
  \multicolumn{3}{cV{2}}{Raw Pointer} &
  \multicolumn{3}{cV{2}}{Unsafe} &
  \multicolumn{3}{cV{2}}{Unsafe Call} &
  \multicolumn{3}{cV{2}}{Unsafe}\bigstrut[t]
  \\
\multicolumn{1}{lV{2}}{} &
  \multicolumn{3}{cV{2}}{Declarations ($\nabla$)} &
  \multicolumn{3}{cV{2}}{Deferences  ($\nabla$)} &
  \multicolumn{3}{cV{2}}{Lines ($\nabla$)} & 
  \multicolumn{3}{cV{2}}{Expressions ($\nabla$)} &
  \multicolumn{3}{cV{2}}{Type Casts ($\nabla$)}
  \bigstrut[b]\\
  \clineB{2-16}{2}
\multicolumn{1}{lV{2}}{} &
  \multicolumn{1}{rV{2}}{\rotatebox{90}{Before~~}} &
  \multicolumn{1}{rV{2}}{{\rotatebox{90}{After}}} &
  \multicolumn{1}{cV{2}}{$\Delta\%$} &
  \multicolumn{1}{rV{2}}{\rotatebox{90}{Before~~}} &
  \multicolumn{1}{rV{2}}{{\rotatebox{90}{After}}} &
  \multicolumn{1}{cV{2}}{$\Delta\%$} &
  \multicolumn{1}{rV{2}}{\rotatebox{90}{Before~~}} &
  \multicolumn{1}{rV{2}}{{\rotatebox{90}{After}}} &
  \multicolumn{1}{cV{2}}{$\Delta\%$} &
  \multicolumn{1}{rV{2}}{\rotatebox{90}{Before~~}} &
  \multicolumn{1}{rV{2}}{{\rotatebox{90}{After}}} &
  \multicolumn{1}{cV{2}}{$\Delta\%$} &
  \multicolumn{1}{rV{2}}{\rotatebox{90}{Before~~}} &
  \multicolumn{1}{rV{2}}{{\rotatebox{90}{After}}} &
  \multicolumn{1}{cV{2}}{$\Delta\%$}
  \bigstrut\\
  \clineB{2-16}{2}
 &
  \multicolumn{1}{l}{} &
  \multicolumn{1}{l}{} &
  \multicolumn{1}{l}{} &
  \multicolumn{1}{l}{} &
  \multicolumn{1}{l}{} &
  \multicolumn{1}{l}{} &
  \multicolumn{1}{l}{} &
  \multicolumn{1}{l}{} &
  \multicolumn{1}{l}{} &
  \multicolumn{1}{l}{} &
  \multicolumn{1}{l}{} &
  \multicolumn{1}{l}{} &
  \multicolumn{1}{l}{} &
  \multicolumn{1}{l}{}
  \bigstrut\\[-1.33em] \hlineB{2}
\multicolumn{1}{V{2}lV{2}}{split}
& \multicolumn{1}{rV{2}}{141}
& \multicolumn{1}{rV{2}}{130}
& \multicolumn{1}{rV{2}}{\cellcolor{blue!12}8}
& \multicolumn{1}{rV{2}}{463}
& \multicolumn{1}{rV{2}}{405}
& \multicolumn{1}{rV{2}}{\cellcolor{blue!12}13}
& \multicolumn{1}{rV{2}}{6836}
& \multicolumn{1}{rV{2}}{6134}
& \multicolumn{1}{rV{2}}{\cellcolor{blue!12}10}
& \multicolumn{1}{rV{2}}{1405}
& \multicolumn{1}{rV{2}}{1334}
& \multicolumn{1}{rV{2}}{\cellcolor{blue!12}5}
& \multicolumn{1}{rV{2}}{5400}
& \multicolumn{1}{rV{2}}{4427}
& \multicolumn{1}{rV{2}}{\cellcolor{blue!12}18}
\bigstrut\\
\multicolumn{1}{V{2}lV{2}}{pwd}
& \multicolumn{1}{rV{2}}{38}
& \multicolumn{1}{rV{2}}{40}
& \multicolumn{1}{rV{2}}{\cellcolor{blue!12}-5}
& \multicolumn{1}{rV{2}}{85}
& \multicolumn{1}{rV{2}}{74}
& \multicolumn{1}{rV{2}}{\cellcolor{blue!12}13}
& \multicolumn{1}{rV{2}}{704}
& \multicolumn{1}{rV{2}}{565}
& \multicolumn{1}{rV{2}}{\cellcolor{blue!30}\textbf{20}}
& \multicolumn{1}{rV{2}}{217}
& \multicolumn{1}{rV{2}}{223}
& \multicolumn{1}{rV{2}}{\cellcolor{blue!12}-3}
& \multicolumn{1}{rV{2}}{646}
& \multicolumn{1}{rV{2}}{292}
& \multicolumn{1}{rV{2}}{\cellcolor{blue!30}\textbf{55}}
\bigstrut\\
\multicolumn{1}{V{2}lV{2}}{cat}
& \multicolumn{1}{rV{2}}{93}
& \multicolumn{1}{rV{2}}{56}
& \multicolumn{1}{rV{2}}{\cellcolor{blue!30}\textbf{40}}
& \multicolumn{1}{rV{2}}{242}
& \multicolumn{1}{rV{2}}{200}
& \multicolumn{1}{rV{2}}{\cellcolor{blue!30}\textbf{17}}
& \multicolumn{1}{rV{2}}{3508}
& \multicolumn{1}{rV{2}}{2828}
& \multicolumn{1}{rV{2}}{\cellcolor{blue!12}19}
& \multicolumn{1}{rV{2}}{659}
& \multicolumn{1}{rV{2}}{572}
& \multicolumn{1}{rV{2}}{\cellcolor{blue!30}\textbf{13}}
& \multicolumn{1}{rV{2}}{2625}
& \multicolumn{1}{rV{2}}{1587}
& \multicolumn{1}{rV{2}}{\cellcolor{blue!12}40}
\bigstrut\\
\multicolumn{1}{V{2}lV{2}}{truncate}
& \multicolumn{1}{rV{2}}{68}
& \multicolumn{1}{rV{2}}{66}
& \multicolumn{1}{rV{2}}{\cellcolor{blue!12}3}
& \multicolumn{1}{rV{2}}{241}
& \multicolumn{1}{rV{2}}{211}
& \multicolumn{1}{rV{2}}{\cellcolor{blue!12}12}
& \multicolumn{1}{rV{2}}{4050}
& \multicolumn{1}{rV{2}}{3636}
& \multicolumn{1}{rV{2}}{\cellcolor{blue!12}10}
& \multicolumn{1}{rV{2}}{772}
& \multicolumn{1}{rV{2}}{676}
& \multicolumn{1}{rV{2}}{\cellcolor{blue!12}12}
& \multicolumn{1}{rV{2}}{4231}
& \multicolumn{1}{rV{2}}{3332}
& \multicolumn{1}{rV{2}}{\cellcolor{blue!12}21}
\bigstrut\\
\multicolumn{1}{V{2}lV{2}}{uniq}
& \multicolumn{1}{rV{2}}{85}
& \multicolumn{1}{rV{2}}{71}
& \multicolumn{1}{rV{2}}{\cellcolor{blue!12}16}
& \multicolumn{1}{rV{2}}{156}
& \multicolumn{1}{rV{2}}{144}
& \multicolumn{1}{rV{2}}{\cellcolor{blue!12}8}
& \multicolumn{1}{rV{2}}{3027}
& \multicolumn{1}{rV{2}}{2636}
& \multicolumn{1}{rV{2}}{\cellcolor{blue!12}13}
& \multicolumn{1}{rV{2}}{569}
& \multicolumn{1}{rV{2}}{516}
& \multicolumn{1}{rV{2}}{\cellcolor{blue!12}9}
& \multicolumn{1}{rV{2}}{3377}
& \multicolumn{1}{rV{2}}{2917}
& \multicolumn{1}{rV{2}}{\cellcolor{blue!12}14}
\bigstrut\\
\multicolumn{1}{V{2}lV{2}}{tail}
& \multicolumn{1}{rV{2}}{147}
& \multicolumn{1}{rV{2}}{125}
& \multicolumn{1}{rV{2}}{\cellcolor{blue!12}15}
& \multicolumn{1}{rV{2}}{568}
& \multicolumn{1}{rV{2}}{475}
& \multicolumn{1}{rV{2}}{\cellcolor{blue!12}16}
& \multicolumn{1}{rV{2}}{7203}
& \multicolumn{1}{rV{2}}{6191}
& \multicolumn{1}{rV{2}}{\cellcolor{blue!12}14}
& \multicolumn{1}{rV{2}}{1559}
& \multicolumn{1}{rV{2}}{1514}
& \multicolumn{1}{rV{2}}{\cellcolor{blue!12}3}
& \multicolumn{1}{rV{2}}{5619}
& \multicolumn{1}{rV{2}}{4129}
& \multicolumn{1}{rV{2}}{\cellcolor{blue!12}27}
\bigstrut\\
\multicolumn{1}{V{2}lV{2}}{head}
& \multicolumn{1}{rV{2}}{87}
& \multicolumn{1}{rV{2}}{75}
& \multicolumn{1}{rV{2}}{\cellcolor{blue!12}14}
& \multicolumn{1}{rV{2}}{340}
& \multicolumn{1}{rV{2}}{297}
& \multicolumn{1}{rV{2}}{\cellcolor{blue!12}13}
& \multicolumn{1}{rV{2}}{3724}
& \multicolumn{1}{rV{2}}{3160}
& \multicolumn{1}{rV{2}}{\cellcolor{blue!12}15}
& \multicolumn{1}{rV{2}}{903}
& \multicolumn{1}{rV{2}}{801}
& \multicolumn{1}{rV{2}}{\cellcolor{blue!12}11}
& \multicolumn{1}{rV{2}}{3193}
& \multicolumn{1}{rV{2}}{2355}
& \multicolumn{1}{rV{2}}{\cellcolor{blue!12}26}
\bigstrut\\
\hlineB{2}
 &
  \multicolumn{1}{l}{} &
  \multicolumn{1}{l}{} &
  \multicolumn{1}{l}{} &
  \multicolumn{1}{l}{} &
  \multicolumn{1}{l}{} &
  \multicolumn{1}{l}{} &
  \multicolumn{1}{l}{} &
  \multicolumn{1}{l}{} &
  \multicolumn{1}{l}{} &
  \multicolumn{1}{l}{} &
  \multicolumn{1}{l}{} &
  \multicolumn{1}{l}{} &
  \multicolumn{1}{l}{} &
  \multicolumn{1}{l}{}
  \bigstrut\\[-1.33em]
\hlineB{2}
\multicolumn{1}{V{2}lV{2}}{avg. $\Delta\%$}
& \multicolumn{1}{rV{2}}{}
& \multicolumn{1}{rV{2}}{}
& \multicolumn{1}{rV{2}}{\cellcolor{blue!12}13}
& \multicolumn{1}{rV{2}}{}
& \multicolumn{1}{rV{2}}{}
& \multicolumn{1}{rV{2}}{\cellcolor{blue!12}13}
& \multicolumn{1}{rV{2}}{}
& \multicolumn{1}{rV{2}}{}
& \multicolumn{1}{rV{2}}{\cellcolor{blue!12}14}
& \multicolumn{1}{rV{2}}{}
& \multicolumn{1}{rV{2}}{}
& \multicolumn{1}{rV{2}}{\cellcolor{blue!12}7}
& \multicolumn{1}{rV{2}}{}
& \multicolumn{1}{rV{2}}{}
& \multicolumn{1}{rV{2}}{\cellcolor{blue!12}29}
\bigstrut\\
\hlineB{2}
\end{tabular}%
}
\vspace{0.7em}
\subcaption{\rebuttal{[Reviewer 2 - Test coverage]}{Our metrics for measuring safety of Rust code, computed across all applications, filtered to include covered functions only.}}
\end{minipage}

\end{table*}

\noindent\textit{\textbf{Results:}} \Cref{tab:coreutils} tabulates
the effectiveness of translating CoreUtils projects into safer Rust using \tool. In the following, we discuss the key findings in three key metrics: raw pointer declarations, raw pointer dereferences, and line of unsafe code.    

\be
\item \textit{Raw Pointer Declarations.} We observe an average reduction of raw pointer declarations by $\mathbf{25\%}$ with notable reductions across all projects. The most significant improvement is observed in \texttt{cat}, which shows a $\mathbf{38\%}$ reduction in raw pointer declarations. \rebuttal{}{Unexpectedly, we see an \textit{increase} in this metric for \TT{pwd} in the filtered seting. This does \textit{not} signify a decrease in safety of the program; rather, it is due to a quirk of the way this metric is computed. The metric counts only variable \textit{bindings} as declarations. So \TT{foo("abc" as const *u8)} would not be counted as a raw pointer declaration, but \TT{let arg = "abc" as const *u8; foo(arg)} would be.}
\item \textit{Raw Pointer Dereferences.} Unsafe raw pointer dereferences are reduced on average by $\mathbf{22\%}$, demonstrating the tool's effectiveness in minimizing unsafe manipulations. The project \texttt{uniq} shows the largest decrease at $\mathbf{27\%}$, indicating a significant impact on pointer-heavy codebases.
\ee


We also observed a notable reduction in unsafe code across all CoreUtils projects after translation to safer Rust.
\be
\item \textit{Lines of unsafe code.} Lines of unsafe code decrease by an average of $\mathbf{24\%}$ across all projects, reflecting a significant enhancement in code safety and idiomaticity. \texttt{uniq} shows the largest reduction of $\mathbf{28\%}$, highlighting \tool's ability to minimize unsafe code.
\item \textit{Unsafe Call Expression:} Unsafe function calls decrease, but by a relatively smaller amount compared to the other metrics - an average of $\mathbf{8\%}$ reduction. The most substantial decrease is in \texttt{head}, from 1378 to 1189 casts (by $\approx\mathbf{14}\%$). The sub-par overall performance is due to the reliance on FFI calls, which is discussed further below.

\item \textit{Unsafe Type Cast:} Unsafe type casts decrease by an average of 30\%. The most substantial decrease is in \texttt{tail}, from 5869 to 3739 casts (by $\approx\mathbf{36}\%$).


\ee

\begin{figure*}
\centering
\resizebox{0.75\linewidth}{!}{%
    \begin{minipage}{0.48\linewidth}
    \begin{minipage}{\linewidth}
    \includegraphics[width=\linewidth]{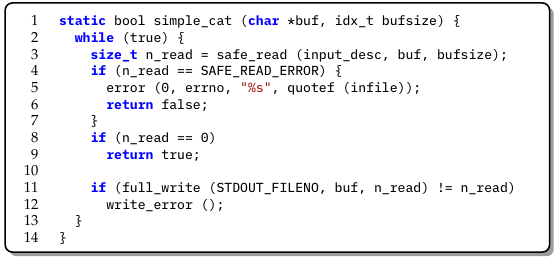}
    \subcaption{Original C implementation}
    \label{subfig:rq1a}
    \end{minipage}\vspace{0.55em}
    \begin{minipage}{\linewidth}
    \includegraphics[width=\linewidth]{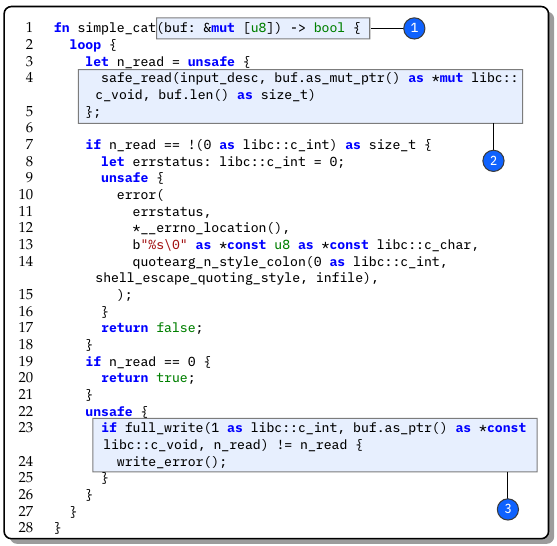}
    \subcaption{\tool output}
    \label{subfig:rq1b}
    \end{minipage}
    \end{minipage}~%
    \begin{minipage}{0.4975\linewidth}
    \includegraphics[width=\linewidth]{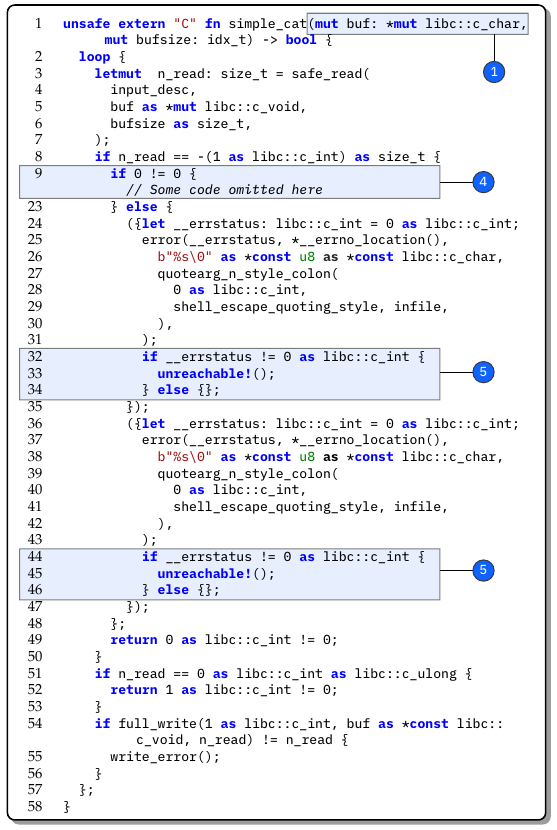}
    \subcaption{C2Rust translation}
    \label{subfig:rq1c}
    \end{minipage}%
}
\caption{C code versus C2Rust code versus LLM-translated safer Rust, for the \TT{simple\_cat} function from the \TT{cat} program.}
\label{fig:manual_analysis}
\end{figure*}
    
\noindent\textbf{Discussion:} \Cref{fig:manual_analysis} illustrates a function using the original C code (\cref{subfig:rq1a}), C2Rust translation (\cref{subfig:rq1b}), and \tool's safer output (\cref{subfig:rq1c}) for comparison. We make the following observations:
\be
\item[\textbf{1)}] \textbf{\tool uses idiomatic and native Rust types.} Non-native types in the C2Rust code (such as \TT{libc::c\_char}, see \circled{1} in \cref{subfig:rq1b}) are replaced with their native Rust types (i.e., \TT{\&mut [u8]} as seen in \circled{1}, \cref{subfig:rq1b}). Note that in some cases these pointers are cast back into their raw pointer-type when passed as an argument to other functions, e.g., \circled{2} and \circled{3} in \cref{subfig:rq1b}.

\item[\textbf{2)}] \textbf{\tool continues to make use of FFI calls.} \tool accomplishes certain core functionalities through calls to C API functions via foreign function interfaces (FFIs), e.g., \TT{safe\_read} and \TT{error} (see \circled{2} and \circled{3} in \cref{subfig:rq1b}). This limits the extent of code safety since the arguments and return types of these functions still use raw pointer types. These may only be eliminated when all the external functions are replaced with safe Rust equivalents; this is beyond the scope of this work.

\item[\textbf{3)}] \textbf{\tool eliminates nonstandard control flow structures and dead code.} Rust code generated by the C2Rust transpiler generates a large number of unreachable branches in the C2Rust code. It makes explicit calls to branch conditions that cannot be executed (e.g., \TT{if 0 != 0} as seen in \circled{4}, \cref{subfig:rq1c}). C2Rust also uses certain macros such as \TT{unreachable!()} (as seen in \circled{5}, \cref{subfig:rq1c}). \tool effectively eliminates these dead code fragments and produces more concise and idiomatic Rust code.
\ee
\begin{result}
\tool reduces unsafe constructs in Coreutils, cutting raw pointers by 38\% and dereferences by 27\%. Unsafe lines, type casts and calls decrease by 24\%, 8\% and 24\%, enhancing safety and idiomaticity.
\end{result}

\begin{table*}[!t]
\centering
\caption{A comparative evaluation of \tool and other state-of-the-art methods against the baseline (C2Rust). Lower values are preferred. Cells in \colorbox{blue!12}{blue} show the greatest improvement on that metric for that project. \tool presents the most significant improvement as indicated by the best Win/Tie/Loss ratio. Laertes and \textsc{crown} actually cause an \textit{increase} in the amount of unsafe code in several translations, as indicated by the negative values under the $\%\Delta$ columns.}
\label{tab:rq2}
\resizebox{0.7\linewidth}{!}{
\begin{tabular}{ccrrrrrrrrrr}
\clineB{3-12}{2}
\multicolumn{2}{cV{2}}{\multirow{2}{*}{}} &
  \multicolumn{2}{cV{2}}{\begin{tabular}[c]{@{}c@{}}Raw Pointer\bigstrut[t]\\ Declaration\end{tabular}} &
  \multicolumn{2}{cV{2}}{\begin{tabular}[c]{@{}c@{}}Raw Pointer\bigstrut[t]\\ Dereferences\end{tabular}} &
  \multicolumn{2}{cV{2}}{\begin{tabular}[c]{@{}c@{}}Lines of\bigstrut[t]\\Unsafe Code\end{tabular}} &
  \multicolumn{2}{cV{2}}{\begin{tabular}[c]{@{}c@{}}Unsafe Call\bigstrut[t]\\Expressions\end{tabular}} &
  \multicolumn{2}{cV{2}}{\begin{tabular}[c]{@{}c@{}}Unsafe\bigstrut[t]\\Type Casts\end{tabular}} \bigstrut\\ \clineB{3-12}{2}
\multicolumn{2}{cV{2}}{} &
  ~ &
  \multicolumn{1}{rV{2}}{\%$\Delta$} &
  ~ &
  \multicolumn{1}{rV{2}}{\%$\Delta$} &
  ~ &
  \multicolumn{1}{rV{2}}{\%$\Delta$} &
  ~ &
  \multicolumn{1}{rV{2}}{\%$\Delta$} &
  ~ &
  \multicolumn{1}{rV{2}}{\%$\Delta$} \bigstrut[t]\\ \clineB{3-12}{2}
\multicolumn{1}{l}{} &
  \multicolumn{1}{l}{} &
  \multicolumn{1}{l}{} &
  \multicolumn{1}{l}{} &
  \multicolumn{1}{l}{} &
  \multicolumn{1}{l}{} &
  \multicolumn{1}{l}{} &
  \multicolumn{1}{l}{} &
  \multicolumn{1}{l}{} &
  \multicolumn{1}{l}{} &
  \multicolumn{1}{l}{} &
  \multicolumn{1}{l}{} \bigstrut\\[-1em] \hlineB{2}
\multicolumn{1}{V{2}cV{2}}{\multirow{5.5}{*}{bzip2}} & \multicolumn{1}{cV{2}}{Baseline} & 227 & \multicolumn{1}{rV{2}}{~} & 3764 & \multicolumn{1}{rV{2}}{~} & 12594 & \multicolumn{1}{rV{2}}{~} & 2580 & \multicolumn{1}{rV{2}}{~} & 7614 & \multicolumn{1}{rV{2}}{~} \bigstrut \\
\multicolumn{1}{V{2}cV{2}}{} & \multicolumn{1}{cV{2}}{Laertes}
& 206 & \multicolumn{1}{rV{2}}{9}& 3699 & \multicolumn{1}{rV{2}}{2}& 13115 & \multicolumn{1}{rV{2}}{-4}& 2788 & \multicolumn{1}{rV{2}}{-8}& 8699 & \multicolumn{1}{rV{2}}{-14} \bigstrut\\
\multicolumn{1}{V{2}cV{2}}{} & \multicolumn{1}{cV{2}}{\textsc{crown}}
& \cellcolor{blue!12}{194} & \multicolumn{1}{rV{2}}{\cellcolor{blue!12}{15}}& 3674 & \multicolumn{1}{rV{2}}{2}& 12592 & \multicolumn{1}{rV{2}}{0}& 2929 & \multicolumn{1}{rV{2}}{-14}& 7672 & \multicolumn{1}{rV{2}}{-1} \bigstrut\\
\multicolumn{1}{V{2}cV{2}}{} & \multicolumn{1}{cV{2}}{\tool}
& 199 & \multicolumn{1}{rV{2}}{12}& \cellcolor{blue!12}{2691} & \multicolumn{1}{rV{2}}{\cellcolor{blue!12}{29}}& \cellcolor{blue!12}{10328} & \multicolumn{1}{rV{2}}{\cellcolor{blue!12}{18}}& \cellcolor{blue!12}{2455} & \multicolumn{1}{rV{2}}{\cellcolor{blue!12}{5}}& \cellcolor{blue!12}{4097} & \multicolumn{1}{rV{2}}{\cellcolor{blue!12}{46}} \bigstrut\\
\hlineB{2}
\multicolumn{1}{V{2}cV{2}}{\multirow{5.5}{*}{genann}} & \multicolumn{1}{cV{2}}{Baseline} & 73 & \multicolumn{1}{rV{2}}{~} & 339 & \multicolumn{1}{rV{2}}{~} & 1634 & \multicolumn{1}{rV{2}}{~} & 579 & \multicolumn{1}{rV{2}}{~} & 1319 & \multicolumn{1}{rV{2}}{~} \bigstrut \\
\multicolumn{1}{V{2}cV{2}}{} & \multicolumn{1}{cV{2}}{Laertes}
& 73 & \multicolumn{1}{rV{2}}{0}& 339 & \multicolumn{1}{rV{2}}{0}& 1650 & \multicolumn{1}{rV{2}}{-1}& 581 & \multicolumn{1}{rV{2}}{-0}& 1329 & \multicolumn{1}{rV{2}}{-1} \bigstrut\\
\multicolumn{1}{V{2}cV{2}}{} & \multicolumn{1}{cV{2}}{\textsc{crown}}
& 71 & \multicolumn{1}{rV{2}}{3}& 317 & \multicolumn{1}{rV{2}}{6}& 1775 & \multicolumn{1}{rV{2}}{-9}& 643 & \multicolumn{1}{rV{2}}{-11}& 1317 & \multicolumn{1}{rV{2}}{0} \bigstrut\\
\multicolumn{1}{V{2}cV{2}}{} & \multicolumn{1}{cV{2}}{\tool}
& \cellcolor{blue!12}{63} & \multicolumn{1}{rV{2}}{\cellcolor{blue!12}{14}}& \cellcolor{blue!12}{273} & \multicolumn{1}{rV{2}}{\cellcolor{blue!12}{19}}& \cellcolor{blue!12}{994} & \multicolumn{1}{rV{2}}{\cellcolor{blue!12}{39}}& \cellcolor{blue!12}{578} & \multicolumn{1}{rV{2}}{\cellcolor{blue!12}{0}}& \cellcolor{blue!12}{362} & \multicolumn{1}{rV{2}}{\cellcolor{blue!12}{73}} \bigstrut\\
\hlineB{2}
\multicolumn{1}{V{2}cV{2}}{\multirow{5.5}{*}{lil}} & \multicolumn{1}{cV{2}}{Baseline} & 438 & \multicolumn{1}{rV{2}}{~} & 1668 & \multicolumn{1}{rV{2}}{~} & 4852 & \multicolumn{1}{rV{2}}{~} & 1809 & \multicolumn{1}{rV{2}}{~} & 2788 & \multicolumn{1}{rV{2}}{~} \bigstrut \\
\multicolumn{1}{V{2}cV{2}}{} & \multicolumn{1}{cV{2}}{Laertes}
& 430 & \multicolumn{1}{rV{2}}{2}& 1658 & \multicolumn{1}{rV{2}}{1}& 4645 & \multicolumn{1}{rV{2}}{4}& 1844 & \multicolumn{1}{rV{2}}{-2}& 2727 & \multicolumn{1}{rV{2}}{2} \bigstrut\\
\multicolumn{1}{V{2}cV{2}}{} & \multicolumn{1}{cV{2}}{\textsc{crown}}
& \cellcolor{blue!12}{355} & \multicolumn{1}{rV{2}}{\cellcolor{blue!12}{19}}& \cellcolor{blue!12}{756} & \multicolumn{1}{rV{2}}{\cellcolor{blue!12}{55}}& 5073 & \multicolumn{1}{rV{2}}{-5}& 3383 & \multicolumn{1}{rV{2}}{-87}& 2523 & \multicolumn{1}{rV{2}}{10} \bigstrut\\
\multicolumn{1}{V{2}cV{2}}{} & \multicolumn{1}{cV{2}}{\tool}
& 428 & \multicolumn{1}{rV{2}}{2}& 1449 & \multicolumn{1}{rV{2}}{13}& \cellcolor{blue!12}{4582} & \multicolumn{1}{rV{2}}{\cellcolor{blue!12}{6}}& \cellcolor{blue!12}{1843} & \multicolumn{1}{rV{2}}{\cellcolor{blue!12}{-2}}& \cellcolor{blue!12}{2111} & \multicolumn{1}{rV{2}}{\cellcolor{blue!12}{24}} \bigstrut\\
\hlineB{2}
\multicolumn{1}{V{2}cV{2}}{\multirow{5.5}{*}{tulipindicators}} & \multicolumn{1}{cV{2}}{Baseline} & 866 & \multicolumn{1}{rV{2}}{~} & 1847 & \multicolumn{1}{rV{2}}{~} & 17671 & \multicolumn{1}{rV{2}}{~} & 2595 & \multicolumn{1}{rV{2}}{~} & 12983 & \multicolumn{1}{rV{2}}{~} \bigstrut \\
\multicolumn{1}{V{2}cV{2}}{} & \multicolumn{1}{cV{2}}{Laertes}
& 866 & \multicolumn{1}{rV{2}}{0}& 1847 & \multicolumn{1}{rV{2}}{0}& 25594 & \multicolumn{1}{rV{2}}{-45}& 2595 & \multicolumn{1}{rV{2}}{0}& 13050 & \multicolumn{1}{rV{2}}{-1} \bigstrut\\
\multicolumn{1}{V{2}cV{2}}{} & \multicolumn{1}{cV{2}}{\textsc{crown}}
& 865 & \multicolumn{1}{rV{2}}{0}& 1847 & \multicolumn{1}{rV{2}}{0}& 17769 & \multicolumn{1}{rV{2}}{-1}& \cellcolor{blue!12}{2419} & \multicolumn{1}{rV{2}}{\cellcolor{blue!12}{7}}& 13195 & \multicolumn{1}{rV{2}}{-2} \bigstrut\\
\multicolumn{1}{V{2}cV{2}}{} & \multicolumn{1}{cV{2}}{\tool}
& \cellcolor{blue!12}{860} & \multicolumn{1}{rV{2}}{\cellcolor{blue!12}{1}}& \cellcolor{blue!12}{1625} & \multicolumn{1}{rV{2}}{\cellcolor{blue!12}{12}}& \cellcolor{blue!12}{17314} & \multicolumn{1}{rV{2}}{\cellcolor{blue!12}{2}}& 2638 & \multicolumn{1}{rV{2}}{-2}& \cellcolor{blue!12}{12274} & \multicolumn{1}{rV{2}}{\cellcolor{blue!12}{5}} \bigstrut\\
\hlineB{2}
\multicolumn{1}{V{2}cV{2}}{\multirow{4}{*}{urlparser}} & \multicolumn{1}{cV{2}}{Baseline} & 79 & \multicolumn{1}{rV{2}}{~} & 60 & \multicolumn{1}{rV{2}}{~} & 845 & \multicolumn{1}{rV{2}}{~} & 447 & \multicolumn{1}{rV{2}}{~} & 648 & \multicolumn{1}{rV{2}}{~} \bigstrut \\
\multicolumn{1}{V{2}cV{2}}{} & \multicolumn{1}{cV{2}}{Laertes}
& \cellcolor{blue!12}{74} & \multicolumn{1}{rV{2}}{\cellcolor{blue!12}{6}}& \cellcolor{blue!12}{2} & \multicolumn{1}{rV{2}}{\cellcolor{blue!12}{97}}& 1002 & \multicolumn{1}{rV{2}}{-19}& 572 & \multicolumn{1}{rV{2}}{-28}& 1167 & \multicolumn{1}{rV{2}}{-80} \bigstrut\\
\multicolumn{1}{V{2}cV{2}}{} & \multicolumn{1}{cV{2}}{\tool}
& 81 & \multicolumn{1}{rV{2}}{-3}& 51 & \multicolumn{1}{rV{2}}{15}& \cellcolor{blue!12}{702} & \multicolumn{1}{rV{2}}{\cellcolor{blue!12}{17}}& \cellcolor{blue!12}{450} & \multicolumn{1}{rV{2}}{\cellcolor{blue!12}{-1}}& \cellcolor{blue!12}{439} & \multicolumn{1}{rV{2}}{\cellcolor{blue!12}{32}} \bigstrut\\
\hlineB{2}
\multicolumn{1}{V{2}cV{2}}{\multirow{4}{*}{grabc}} & \multicolumn{1}{cV{2}}{Baseline} & 13 & \multicolumn{1}{rV{2}}{~} & 21 & \multicolumn{1}{rV{2}}{~} & 214 & \multicolumn{1}{rV{2}}{~} & 48 & \multicolumn{1}{rV{2}}{~} & 123 & \multicolumn{1}{rV{2}}{~} \bigstrut \\
\multicolumn{1}{V{2}cV{2}}{} & \multicolumn{1}{cV{2}}{Laertes}
& 8 & \multicolumn{1}{rV{2}}{38}& 12 & \multicolumn{1}{rV{2}}{43}& 213 & \multicolumn{1}{rV{2}}{0}& 71 & \multicolumn{1}{rV{2}}{-48}& 122 & \multicolumn{1}{rV{2}}{1} \bigstrut\\
\multicolumn{1}{V{2}cV{2}}{} & \multicolumn{1}{cV{2}}{\tool}
& \cellcolor{blue!12}{7} & \multicolumn{1}{rV{2}}{\cellcolor{blue!12}{46}}& \cellcolor{blue!12}{11} & \multicolumn{1}{rV{2}}{\cellcolor{blue!12}{48}}& \cellcolor{blue!12}{84} & \multicolumn{1}{rV{2}}{\cellcolor{blue!12}{61}}& \cellcolor{blue!12}{33} & \multicolumn{1}{rV{2}}{\cellcolor{blue!12}{31}}& \cellcolor{blue!12}{10} & \multicolumn{1}{rV{2}}{\cellcolor{blue!12}{92}} \bigstrut\\
\hlineB{2}
\multicolumn{1}{V{2}cV{2}}{\multirow{4}{*}{optipng}} & \multicolumn{1}{cV{2}}{Baseline} & 1407 & \multicolumn{1}{rV{2}}{~} & 6062 & \multicolumn{1}{rV{2}}{~} & 82616 & \multicolumn{1}{rV{2}}{~} & 7854 & \multicolumn{1}{rV{2}}{~} & 32913 & \multicolumn{1}{rV{2}}{~} \bigstrut \\
\multicolumn{1}{V{2}cV{2}}{} & \multicolumn{1}{cV{2}}{Laertes}
& 1235 & \multicolumn{1}{rV{2}}{12}& 5552 & \multicolumn{1}{rV{2}}{8}& 93696 & \multicolumn{1}{rV{2}}{-13}& 9189 & \multicolumn{1}{rV{2}}{-17}& 46450 & \multicolumn{1}{rV{2}}{-41} \bigstrut\\
\multicolumn{1}{V{2}cV{2}}{} & \multicolumn{1}{cV{2}}{\tool}
& \cellcolor{blue!12}{1197} & \multicolumn{1}{rV{2}}{\cellcolor{blue!12}{15}}& \cellcolor{blue!12}{5203} & \multicolumn{1}{rV{2}}{\cellcolor{blue!12}{14}}& \cellcolor{blue!12}{64013} & \multicolumn{1}{rV{2}}{\cellcolor{blue!12}{23}}& \cellcolor{blue!12}{7354} & \multicolumn{1}{rV{2}}{\cellcolor{blue!12}{6}}& \cellcolor{blue!12}{25061} & \multicolumn{1}{rV{2}}{\cellcolor{blue!12}{24}} \bigstrut\\
\hlineB{2}
\multicolumn{1}{V{2}cV{2}}{\multirow{4}{*}{qsort}} & \multicolumn{1}{cV{2}}{Baseline} & 4 & \multicolumn{1}{rV{2}}{~} & 10 & \multicolumn{1}{rV{2}}{~} & 32 & \multicolumn{1}{rV{2}}{~} & 11 & \multicolumn{1}{rV{2}}{~} & 12 & \multicolumn{1}{rV{2}}{~} \bigstrut \\
\multicolumn{1}{V{2}cV{2}}{} & \multicolumn{1}{cV{2}}{Laertes}
& 2 & \multicolumn{1}{rV{2}}{50}& \cellcolor{blue!12}{6} & \multicolumn{1}{rV{2}}{\cellcolor{blue!12}{40}}& 32 & \multicolumn{1}{rV{2}}{0}& 23 & \multicolumn{1}{rV{2}}{-109}& \cellcolor{blue!12}{12} & \multicolumn{1}{rV{2}}{\cellcolor{blue!12}{0}} \bigstrut\\
\multicolumn{1}{V{2}cV{2}}{} & \multicolumn{1}{cV{2}}{\tool}
& \cellcolor{blue!12}{1} & \multicolumn{1}{rV{2}}{\cellcolor{blue!12}{75}}& \cellcolor{blue!12}{6} & \multicolumn{1}{rV{2}}{\cellcolor{blue!12}{40}}& \cellcolor{blue!12}{19} & \multicolumn{1}{rV{2}}{\cellcolor{blue!12}{41}}& \cellcolor{blue!12}{10} & \multicolumn{1}{rV{2}}{\cellcolor{blue!12}{9}}& 13 & \multicolumn{1}{rV{2}}{-8} \bigstrut\\
\hlineB{2}
\multicolumn{1}{V{2}cV{2}}{\multirow{4}{*}{snudown}} & \multicolumn{1}{cV{2}}{Baseline} & 244 & \multicolumn{1}{rV{2}}{~} & 842 & \multicolumn{1}{rV{2}}{~} & 5842 & \multicolumn{1}{rV{2}}{~} & 1587 & \multicolumn{1}{rV{2}}{~} & 3252 & \multicolumn{1}{rV{2}}{~} \bigstrut \\
\multicolumn{1}{V{2}cV{2}}{} & \multicolumn{1}{cV{2}}{Laertes}
& \cellcolor{blue!12}{224} & \multicolumn{1}{rV{2}}{\cellcolor{blue!12}{8}}& 765 & \multicolumn{1}{rV{2}}{9}& 7213 & \multicolumn{1}{rV{2}}{-23}& 1772 & \multicolumn{1}{rV{2}}{-12}& 5940 & \multicolumn{1}{rV{2}}{-83} \bigstrut\\
\multicolumn{1}{V{2}cV{2}}{} & \multicolumn{1}{cV{2}}{\tool}
& 231 & \multicolumn{1}{rV{2}}{5}& \cellcolor{blue!12}{747} & \multicolumn{1}{rV{2}}{\cellcolor{blue!12}{11}}& \cellcolor{blue!12}{5583} & \multicolumn{1}{rV{2}}{\cellcolor{blue!12}{4}}& \cellcolor{blue!12}{1533} & \multicolumn{1}{rV{2}}{\cellcolor{blue!12}{3}}& \cellcolor{blue!12}{2867} & \multicolumn{1}{rV{2}}{\cellcolor{blue!12}{12}} \bigstrut\\
\hlineB{2}
\multicolumn{1}{V{2}cV{2}}{\multirow{4}{*}{xzoom}} & \multicolumn{1}{cV{2}}{Baseline} & 29 & \multicolumn{1}{rV{2}}{~} & 172 & \multicolumn{1}{rV{2}}{~} & 1468 & \multicolumn{1}{rV{2}}{~} & 321 & \multicolumn{1}{rV{2}}{~} & 867 & \multicolumn{1}{rV{2}}{~} \bigstrut \\
\multicolumn{1}{V{2}cV{2}}{} & \multicolumn{1}{cV{2}}{Laertes}
& \cellcolor{blue!12}{29} & \multicolumn{1}{rV{2}}{\cellcolor{blue!12}{0}}& 172 & \multicolumn{1}{rV{2}}{0}& 1530 & \multicolumn{1}{rV{2}}{-4}& \cellcolor{blue!12}{321} & \multicolumn{1}{rV{2}}{\cellcolor{blue!12}{0}}& 897 & \multicolumn{1}{rV{2}}{-3} \bigstrut\\
\multicolumn{1}{V{2}cV{2}}{} & \multicolumn{1}{cV{2}}{\tool}
& \cellcolor{blue!12}{29} & \multicolumn{1}{rV{2}}{\cellcolor{blue!12}{0}}& \cellcolor{blue!12}{168} & \multicolumn{1}{rV{2}}{\cellcolor{blue!12}{2}}& \cellcolor{blue!12}{1467} & \multicolumn{1}{rV{2}}{\cellcolor{blue!12}{0}}& 325 & \multicolumn{1}{rV{2}}{-1}& \cellcolor{blue!12}{631} & \multicolumn{1}{rV{2}}{\cellcolor{blue!12}{27}} \bigstrut\\
\hlineB{3}
    \multicolumn{1}{l}{} &
  \multicolumn{1}{l}{} &
  \multicolumn{1}{l}{} &
  \multicolumn{1}{l}{} &
  \multicolumn{1}{l}{} &
  \multicolumn{1}{l}{} &
  \multicolumn{1}{l}{} &
  \multicolumn{1}{l}{} &
  \multicolumn{1}{l}{} &
  \multicolumn{1}{l}{} &
  \multicolumn{1}{l}{} &
  \multicolumn{1}{l}{} \bigstrut\\[-1em] \hlineB{3}
\multicolumn{1}{V{3}cV{3}}{\multirow{4}{*}{\begin{tabular}{c}
     \textbf{Aggregate}  \bigstrut\\ \textbf{Win/Tie/Loss}
\end{tabular}}} &
  \multicolumn{1}{cV{3}}{\cellcolor{blue!12}\textbf{\tool}} &
  \multicolumn{2}{cV{3}}{\cellcolor{blue!12}\textbf{5/1/4}} &
  \multicolumn{2}{cV{3}}{\cellcolor{blue!12}\textbf{7/1/2}} &
  \multicolumn{2}{cV{3}}{\cellcolor{blue!12}\textbf{10/0/0}} &
  \multicolumn{2}{cV{3}}{\cellcolor{blue!12}\textbf{8/0/2}} &
  \multicolumn{2}{cV{3}}{\cellcolor{blue!12}\textbf{9/0/1}} \bigstrut\\ \clineB{2-12}{3}
\multicolumn{1}{V{3}cV{3}}{} &
  \multicolumn{1}{cV{3}}{\textsc{{crown}}} &
  \multicolumn{2}{cV{3}}{2/0/8} &
  \multicolumn{2}{cV{3}}{1/0/9} &
  \multicolumn{2}{cV{3}}{0/0/10} &
  \multicolumn{2}{cV{3}}{1/0/9} &
  \multicolumn{2}{cV{3}}{0/0/10} \bigstrut\\ 
\multicolumn{1}{V{3}cV{3}}{} &
  \multicolumn{1}{cV{3}}{{Laertes}} &
    \multicolumn{2}{cV{3}}{1/1/8} &
  \multicolumn{2}{cV{3}}{1/1/8} &
  \multicolumn{2}{cV{3}}{0/0/10} &
  \multicolumn{2}{cV{3}}{1/0/9} &
  \multicolumn{2}{cV{3}}{1/0/9}
  \bigstrut\\ \hlineB{2}
\end{tabular}%
}
\end{table*}
\subsection{Comparison with Existing Tools (RQ2)}

\noindent\textit{\textbf{Motivation:}} This section presents a comparative analysis of safe Rust transformations obtained from \tool, as well as methodologies from contemporary approaches Laertes and \textsc{crown} applied to Laertes benchmark.\footnote{The incompatibility of Laertes with the current version of the Rust compiler precluded its execution on our CoreUtils benchmarks due to it being developed on a deprecated compiler version. Therefore, \tool was instead executed on the 10 Laertes benchmarks}\textsuperscript{,~}\footnote{\textsc{crown} and was executed on 4 of the 10 benchmark projects. In the remaining projects, we encountered failure due to limitations in managing unions and variadic arguments~\cite{zhang2023ownership}.}

\noindent\textit{\textbf{Evaluation:}} We measure heuristics similar to the previous research question: (1) the number of raw pointer declarations, (2) the number of raw pointer dereferences, and (3) the quantity of unsafe lines of code. Furthermore, we extend our analysis to include measurements of (1) unsafe type casts, (2) unsafe call expressions, and (3) unsafe let declarations.

\noindent\textit{\textbf{Results:}} The results are shown in \Cref{tab:rq2}. We observe that \tool is capable of reducing the size of unsafe blocks, doing better than the baselines in most cases.
\bi
\item On the two raw pointer metrics (i.e., raw pointer declarations and raw pointer dereferences), \tool performs as well as or better than the baselines on a majority of the benchmark programs with an aggregate Win/Tie/Loss of 5/1/4 and 7/1/2 for raw pointer declarations and raw pointer dereferences respectively. 
\item \tool demonstrates superior efficacy in reducing lines of unsafe code, achieving a reduction of up to 61\% in grabc. An aggregate Win/Tie/Loss of 10/0/0 shows consistent performance across all applications.
\item We also observed a notable reduction in key unsafe expressions and declarations with \tool in all projects. On the other hand, Laertes and \textsc{crown} actually cause an \textit{increase} in unsafe code in several projects, as evidenced by the negative values of $\%\Delta$ in \Cref{tab:rq2} under lines of unsafe code, unsafe type casts and unsafe calls.
\ei
\rebuttal{[Reviewer 3 - qualitative comparison]}{Section \ref{sec:laertes_comparison} in the Appendix shows an example of the outputs of \tool and Laertes, for qualitative comparison.}
\vspace{0.5em}
\begin{result}
\tool outperforms other approaches with a 39/2/9 aggregate win/tie/loss record over all 5 metrics, enhancing Rust safety and idiomaticity.
\end{result}

\subsection{Ablation of \tool design choices (RQ3)}

In this section, we evaluate the individual effect of some of the components of \tool.

\noindent\textit{\textbf{Evaluation:}} We pick 3 arbitrary baselines from coreutils - \TT{cat}, \TT{uniq} and \TT{head} - and run \tool in two settings:
\be
\item \textbf{No Chunking:} Translating each function as a whole without dividing it into smaller chunks. Our hypothesis is that LLMs do better when the input is not too long, so we would expect this setting to be worse than the original.
\item \textbf{Random Order:} Translating the functions in a random order, instead of following the ordering described in \Cref{sec:ordering}.
\ee

\noindent
\revision{\textbf{Function Length:} We also study the effect of function length on translation. For this study, we consider only those functions which are translated as a single unit, \ie those which are smaller than $L = 150$ lines. We define two quantities: 
\bi
\item \textit{Translation Success Rate:} The percentage of functions that translate correctly, where a correct translation is one that compiles and passes tests.
\item \textit{Number of Attempts:} The number of iterations of compiler or test feedback given to the LLM before it generates a successful translation. This is measured only for functions for which a correct translation is generated within $N$ attempts.
\ei
We measure Translation Success Rate and Number of Attempts over functions of varying length, from $0$ to $L=150$.}

\begin{figure}[ht]
    \centering
    \resizebox{0.8\linewidth}{!}{%
    \begin{subfigure}{\linewidth}
        \centering
        \includegraphics[width=0.8\linewidth]{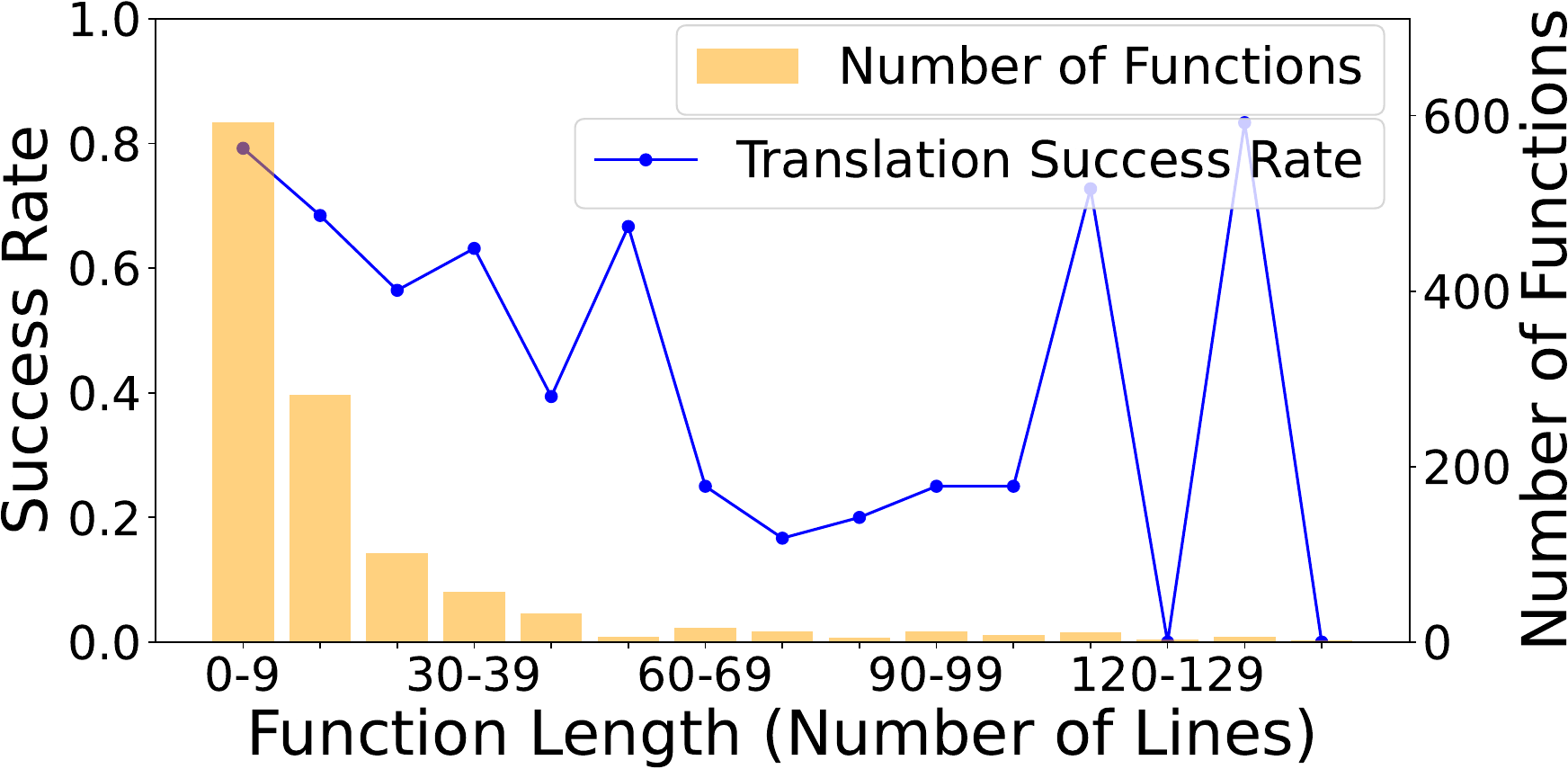}
        \caption{Translation Success Rate versus Function Length}
        \label{fig:success_vs_length}
    \end{subfigure}%
    }
    \resizebox{0.8\linewidth}{!}{%
    \begin{subfigure}{\linewidth}
        \centering
        \includegraphics[width=0.8\linewidth]{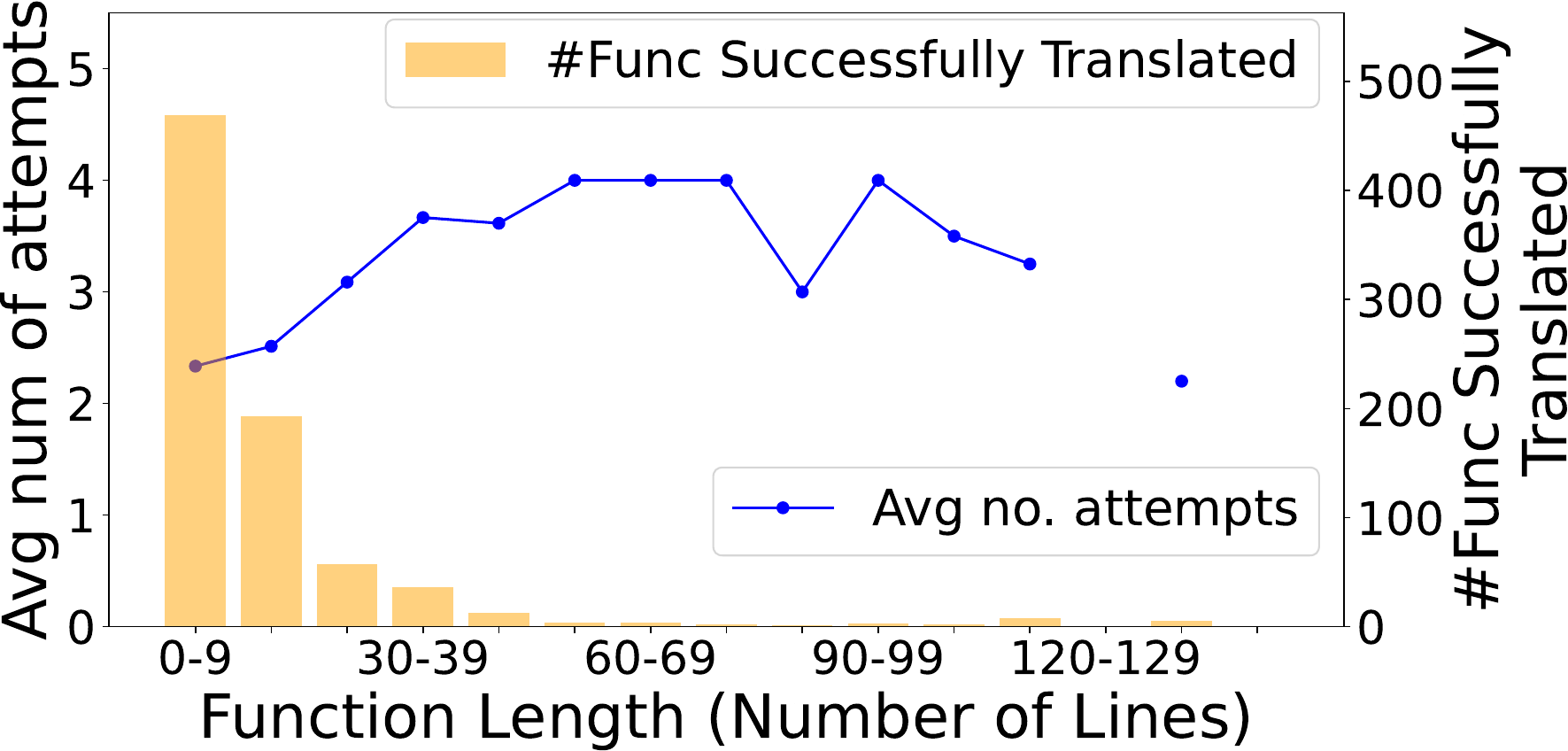}
        \caption{\#Attempts Required to Translate a Function, vs Length}
        \label{fig:attempts_vs_length}
    \end{subfigure}%
    }

    \caption{For our \TT{coreutils} benchmark dataset - a plot of the fraction of successful translations versus function length, and the number of attempts required to translate a function versus function length. This includes only functions that are translated as a single unit, \ie functions that are $\le$ 150 lines.}
    \label{fig:length_plot}
\end{figure}

\noindent\textit{\textbf{Results:}} The results for the component ablation are shown in \Cref{tab:ablation}. There are 3 utilities and 5 metrics, which makes a total of 15 settings. \tool outperforms both ``No Chunking'' as well as ``Random Order'' in 9 out of 15 settings, showing a clear benefit of both design choices.

\Cref{fig:success_vs_length} shows the rate of translation success versus function length, and \Cref{fig:attempts_vs_length} shows the number of attempts needed to translate a function vs its length. If we ignore the outliers on the right, we can see that as the function length increases, the translation success rate decreases and the number of attempts increases.
This provides empirical support for our decision to break up very long functions into smaller chunks for translation.

\begin{table}[t]
    \centering
    \resizebox{\columnwidth}{!}{%
    \begin{tabular}{cV{2}cV{2}cV{2}cV{2}cV{2}cV{2}cV{2}}
\toprule
        \multirow{2}{*}{\textbf{Program}} & \multirow{2}{*}{\textbf{Method}} & \multicolumn{1}{cV{2}}{Raw Ptr} & \multicolumn{1}{cV{2}}{Raw Ptr} & \multicolumn{1}{cV{2}}{Unsafe} & \multicolumn{1}{cV{2}}{Unsafe} & \multicolumn{1}{cV{2}}{Unsafe}\\
         & &  Decls $\downarrow$ & Derefs $\downarrow$ & Lines $\downarrow$ & Calls $\downarrow$ & Casts $\downarrow$ \\
\midrule
\multirow{3}{*}{\textbf{cat}} & Original & 192  & 317 & 5625 & 1038 & 5625 \\
& No Chunking & 137 & 248 & \cellcolor{blue!12}{3638} & 922 & \cellcolor{blue!12}{3638} \\
 & Random Order & 139  & 246  & 4194  & 955  & 4194 \\
 & Our Tool & \cellcolor{blue!12}{120}  & \cellcolor{blue!12}{240}  & 4189  & \cellcolor{blue!12}{912}  & 4189 \\
\midrule
\multirow{3}{*}{\textbf{uniq}} & Original & 227 & 343 & 6066 & 1150 & 6066 \\
& No Chunking & 169 & 252 & \cellcolor{blue!12}{4382} & 1114 & \cellcolor{blue!12}{4382} \\
 & Random Order & \cellcolor{blue!12}{164}  & 258  & 4516  & 1117  & 4516 \\
 & Our Tool & 166  & \cellcolor{blue!12}{250}  & 4397  & \cellcolor{blue!12}{1025}  & 4397 \\
\midrule
\multirow{3}{*}{\textbf{head}} & Original & 192 & 442 & 6245 & 1378 & 6245 \\
& No Chunking & 142 & 360 & 4814 & 1380 & 4814 \\
 & Random Order & 142  & \cellcolor{blue!12}{349}  & 4735  & 1252  & 4735 \\
 & Our Tool & \cellcolor{blue!12}{141}  & 351  & \cellcolor{blue!12}{4699}  & \cellcolor{blue!12}{1189}  & \cellcolor{blue!12}{4699} \\
\bottomrule
    \end{tabular}%
    }
    \caption{Showing the contribution of the components of our method by ablation on three selected benchmarks.}
    \label{tab:ablation}
\end{table}

\section{Related Work}

\textbf{Converting C to unsafe Rust.}
Corrode \cite{corrode} and C2Rust \cite{c2rust} both aim to convert C to functionally equivalent Rust. C2Rust continues to be actively maintained, whereas Corrode has seen little recent development. Both generate non-idiomatic Rust that relies heavily on \TT{unsafe} constructs. Citrus \cite{citrus} similarly provides syntactic mappings from C to Rust, but focuses on being lightweight rather than fully correct: the output may not compile or run without manual revision. It is intended more as a rough starting point than a complete translation pipeline.

\smallskip
\noindent
\textbf{Converting unsafe Rust to safer Rust.}
Because C2Rust code relies extensively on raw pointers, much work has explored automatically replacing them with safer Rust references. Laertes \cite{emre2021translating} was the first to use compiler feedback to infer lifetimes and mutability, rewriting raw pointers when ownership information could be recovered. Their study also analyzed common patterns of \TT{unsafe} in C2Rust output. However, the approach applies only when unsafety arises from missing ownership rather than aliasing or other causes, converting about 8.3\% of raw pointers. A later extension \cite{emre2023aliasing} broadens coverage to more pointer categories and improves conversion rates to about 21\%.
CROWN \cite{zhang2023ownership} also starts from C2Rust output but instead constructs pointer constraints and solves them using an SMT solver. It is more targeted, supporting only mutable, non-array pointers, and converts $\sim$37\% of these to safe references. Other efforts pursue narrower correctness improvements: \citet{hong2024don} replace output-parameter success/failure patterns with \TT{Option}/\TT{Result}, and \cite{hong2024tag} translate C unions into Rust tagged unions. NADER \cite{popescu2021safer} replaces unchecked indexing with checked indexing, allowing a tunable safety–performance tradeoff, but does not address raw pointers or unsafe blocks more generally.

\smallskip
\noindent
\textbf{LLMs for Code Translation.}
Early LLM-based translation work began with TransCoder \cite{roziere2020unsupervised}, using Back Translation for fully unsupervised learning. Later, \citet{ahmad2022summarize} enhanced this with code summarization. \citet{roziere2021leveraging} generate aligned parallel data for supervised training by generate synthetic pairs using another model and verifying them with unit tests. Recent code-focused LLMs have demonstrated strong translation capabilities \cite{pan2024lost}. \citet{yang2024exploring} introduce UniTrans, which generates tests and iteratively repairs translations using execution feedback. \citet{nitin2024spectra} use multi-modal specifications with a self-consistency filter, and \citet{yang2024vert} verify candidate translations against transpiler output using equivalence proofs.
However, these works generally deal with programs that fit in a single model context window, such as competitive programming solutions.

\citet{shiraishi2024context} were the first to attempt translation of arbitrarily large repositories, slicing projects into smaller units, translating each with an LLM, then compiling and repairing errors. Their method does not incorporate test feedback and requires each unit to compile independently, limiting scalability. \citet{ibrahimzada2024repository} translate full Java projects to Python, and \citet{zhang2024scalable} translate Go to Rust with emphasis on avoiding runtime errors. \citet{shetty2024syzygy} perform full C→Rust repository translation by dividing code into units, translating with LLMs, and validating using dynamic analysis with LLM-generated tests. Our approach differs in two key ways: (a) we leverage a transpiler to generate Rust skeletons, enabling end-to-end correctness checking at every iteration; and (b) we decompose code below method granularity, allowing translation of repositories with extremely large functions.
\section{Threats to Validity}

\revision{Although C2Rust makes efforts to ensure functional equivalence between the C and unsafe Rust code, this is not a formal guarantee. \tool{} uses C2Rust as the first step, and so the validity of the entire approach depends on the correctness of C2Rust's transpilation. However, C2Rust merely provides an interface between the original code and translated code that enables the program to be tested end-to-end. It could be replaced with any other such framework, like Rust's Foreign Function Interface (FFI).}

Our metrics are intended to model the safety and naturalness of the code; however it is possible that an improvement in a metric doesn't correlate with improved safety or naturalness of the code. For example, the number of lines of code within unsafe blocks could decrease simply by eliminating line breaks and using semicolons to delimit lines. We therefore assume that the tools we consider do not attempt to game the metrics by performing such manipulations.

The code we generate still uses a significant amount of unsafe code. Some of this is because we do not translate code outside functions, like structure definitions, and interacting with these structures could necessitate the use of some unsafe code. Another observation is that the LLM does not replace C FFI function calls with their safe Rust equivalents; rather, it uses unsafe code to wrap these FFI calls. 
\section{Conclusion}

In this work, we have presented \tool, a system to automatically translate full C repositories to safer Rust. Based on the novel insight that the C2Rust transpiler can be used as an intermediary to bridge between C and Rust, we first convert the entire repository into unsafe Rust. We divide the code repository into smaller ``translation units'', and extract each translation unit along with its context information. We use an LLM to translate each unit to safer Rust, swap in this translation, and propagate the changes to the rest of the code base.
We contribute a dataset of 7 C programs from GNU CoreUtils with their corresponding C2Rust transpiled versions. Each program comes with end-to-end test cases, allowing one to empirically verify the correctness of translation. On this dataset, we show that we reduce raw pointer declarations and dereferences by up to 38\%, and reduce unsafe code by up to 28\%. We also compare our tool with two baselines for making Rust code safer, and observe consistent gains. Future work could look at translating code outside functions, like structure definitions, as well as remapping C FFI calls to safe Rust equivalents.

\typeout{}
\bibliographystyle{plainnat}
\bibliography{main}

\begin{thebibliography}{31}
\providecommand{\natexlab}[1]{#1}
\providecommand{\url}[1]{\texttt{#1}}
\expandafter\ifx\csname urlstyle\endcsname\relax
  \providecommand{\doi}[1]{doi: #1}\else
  \providecommand{\doi}{doi: \begingroup \urlstyle{rm}\Url}\fi

\bibitem[Ahmad et~al.(2021)Ahmad, Tushar, Chakraborty, and Chang]{ahmad2021avatar}
Wasi~Uddin Ahmad, Md~Golam~Rahman Tushar, Saikat Chakraborty, and Kai-Wei Chang.
\newblock Avatar: A parallel corpus for java-python program translation.
\newblock \emph{arXiv preprint arXiv:2108.11590}, 2021.

\bibitem[Ahmad et~al.(2022)Ahmad, Chakraborty, Ray, and Chang]{ahmad2022summarize}
Wasi~Uddin Ahmad, Saikat Chakraborty, Baishakhi Ray, and Kai-Wei Chang.
\newblock Summarize and generate to back-translate: Unsupervised translation of programming languages.
\newblock \emph{arXiv preprint arXiv:2205.11116}, 2022.

\bibitem[Astrauskas et~al.(2020)Astrauskas, Matheja, Poli, M{\"u}ller, and Summers]{astrauskas2020programmers}
Vytautas Astrauskas, Christoph Matheja, Federico Poli, Peter M{\"u}ller, and Alexander~J Summers.
\newblock How do programmers use unsafe rust?
\newblock \emph{Proceedings of the ACM on Programming Languages}, 4\penalty0 (OOPSLA):\penalty0 1--27, 2020.

\bibitem[{Citrus Contributors}(2017)]{citrus}
{Citrus Contributors}.
\newblock citrus.
\newblock \url{https://gitlab.com/citrus-rs/citrus}, 2017.

\bibitem[{Corrode Contributors}(2017)]{corrode}
{Corrode Contributors}.
\newblock corrode.
\newblock \url{https://github.com/jameysharp/corrode}, 2017.

\bibitem[Durumeric et~al.(2014)Durumeric, Li, Kasten, Amann, Beekman, Payer, Weaver, Adrian, Paxson, Bailey, et~al.]{durumeric2014matter}
Zakir Durumeric, Frank Li, James Kasten, Johanna Amann, Jethro Beekman, Mathias Payer, Nicolas Weaver, David Adrian, Vern Paxson, Michael Bailey, et~al.
\newblock The matter of heartbleed.
\newblock In \emph{Proceedings of the 2014 conference on internet measurement conference}, pages 475--488, 2014.

\bibitem[Emre et~al.(2021)Emre, Schroeder, Dewey, and Hardekopf]{emre2021translating}
Mehmet Emre, Ryan Schroeder, Kyle Dewey, and Ben Hardekopf.
\newblock Translating c to safer rust.
\newblock \emph{Proceedings of the ACM on Programming Languages}, 5\penalty0 (OOPSLA):\penalty0 1--29, 2021.

\bibitem[Emre et~al.(2023)Emre, Boyland, Parekh, Schroeder, Dewey, and Hardekopf]{emre2023aliasing}
Mehmet Emre, Peter Boyland, Aesha Parekh, Ryan Schroeder, Kyle Dewey, and Ben Hardekopf.
\newblock Aliasing limits on translating c to safe rust.
\newblock \emph{Proceedings of the ACM on Programming Languages}, 7\penalty0 (OOPSLA1):\penalty0 551--579, 2023.

\bibitem[{Galois}(2018)]{c2rust}
{Galois}.
\newblock {C2Rust}, 08 2018.
\newblock URL \url{https://galois.com/blog/2018/08/c2rust/}.

\bibitem[Hong and Ryu(2024{\natexlab{a}})]{hong2024don}
Jaemin Hong and Sukyoung Ryu.
\newblock Don’t write, but return: Replacing output parameters with algebraic data types in c-to-rust translation.
\newblock \emph{Proc. ACM Program. Lang.}, 8\penalty0 (PLDI), jun 2024{\natexlab{a}}.
\newblock \doi{10.1145/3656406}.
\newblock URL \url{https://doi.org/10.1145/3656406}.

\bibitem[Hong and Ryu(2024{\natexlab{b}})]{hong2024tag}
Jaemin Hong and Sukyoung Ryu.
\newblock To tag, or not to tag: Translating c's unions to rust's tagged unions.
\newblock In \emph{Proceedings of the 39th IEEE/ACM International Conference on Automated Software Engineering}, ASE '24, page 40–52, New York, NY, USA, 2024{\natexlab{b}}. Association for Computing Machinery.
\newblock ISBN 9798400712487.
\newblock \doi{10.1145/3691620.3694985}.
\newblock URL \url{https://doi.org/10.1145/3691620.3694985}.

\bibitem[Ibrahimzada et~al.(2024)Ibrahimzada, Ke, Pawagi, Abid, Pan, Sinha, and Jabbarvand]{ibrahimzada2024repository}
Ali~Reza Ibrahimzada, Kaiyao Ke, Mrigank Pawagi, Muhammad~Salman Abid, Rangeet Pan, Saurabh Sinha, and Reyhaneh Jabbarvand.
\newblock Repository-level compositional code translation and validation.
\newblock \emph{arXiv preprint arXiv:2410.24117}, 2024.

\bibitem[Leveson and Turner(1993)]{leveson1993investigation}
Nancy~G Leveson and Clark~S Turner.
\newblock An investigation of the therac-25 accidents.
\newblock \emph{Computer}, 26\penalty0 (7):\penalty0 18--41, 1993.

\bibitem[Li et~al.(2024)Li, Wang, Li, Saxena, and Kundu]{li2024translating}
Ruishi Li, Bo~Wang, Tianyu Li, Prateek Saxena, and Ashish Kundu.
\newblock Translating c to rust: Lessons from a user study.
\newblock \emph{arXiv preprint arXiv:2411.14174}, 2024.

\bibitem[Liu et~al.(2023)Liu, Xia, Wang, and Zhang]{evalplus}
Jiawei Liu, Chunqiu~Steven Xia, Yuyao Wang, and Lingming Zhang.
\newblock Is your code generated by chat{GPT} really correct? rigorous evaluation of large language models for code generation.
\newblock In \emph{Thirty-seventh Conference on Neural Information Processing Systems}, 2023.
\newblock URL \url{https://openreview.net/forum?id=1qvx610Cu7}.

\bibitem[Nitin et~al.(2024)Nitin, Krishna, and Ray]{nitin2024spectra}
Vikram Nitin, Rahul Krishna, and Baishakhi Ray.
\newblock Spectra: Enhancing the code translation ability of language models by generating multi-modal specifications, 2024.
\newblock URL \url{https://arxiv.org/abs/2405.18574}.

\bibitem[Pan et~al.(2024)Pan, Ibrahimzada, Krishna, Sankar, Wassi, Merler, Sobolev, Pavuluri, Sinha, and Jabbarvand]{pan2024lost}
Rangeet Pan, Ali~Reza Ibrahimzada, Rahul Krishna, Divya Sankar, Lambert~Pouguem Wassi, Michele Merler, Boris Sobolev, Raju Pavuluri, Saurabh Sinha, and Reyhaneh Jabbarvand.
\newblock Lost in translation: A study of bugs introduced by large language models while translating code.
\newblock In \emph{Proceedings of the IEEE/ACM 46th International Conference on Software Engineering}, pages 1--13, 2024.

\bibitem[Peng et~al.(2023)Peng, Quesnelle, Fan, and Shippole]{peng2023yarn}
Bowen Peng, Jeffrey Quesnelle, Honglu Fan, and Enrico Shippole.
\newblock Yarn: Efficient context window extension of large language models.
\newblock \emph{arXiv preprint arXiv:2309.00071}, 2023.

\bibitem[Popescu et~al.(2021)Popescu, Xu, Apostolakis, August, and Levy]{popescu2021safer}
Natalie Popescu, Ziyang Xu, Sotiris Apostolakis, David~I August, and Amit Levy.
\newblock Safer at any speed: automatic context-aware safety enhancement for rust.
\newblock \emph{Proceedings of the ACM on Programming Languages}, 5\penalty0 (OOPSLA):\penalty0 1--23, 2021.

\bibitem[Puri et~al.(2021)Puri, Kung, Janssen, Zhang, Domeniconi, Zolotov, Dolby, Chen, Choudhury, Decker, et~al.]{puri2021codenet}
Ruchir Puri, David~S Kung, Geert Janssen, Wei Zhang, Giacomo Domeniconi, Vladimir Zolotov, Julian Dolby, Jie Chen, Mihir Choudhury, Lindsey Decker, et~al.
\newblock Codenet: A large-scale ai for code dataset for learning a diversity of coding tasks.
\newblock \emph{arXiv preprint arXiv:2105.12655}, 2021.

\bibitem[Roziere et~al.(2020)Roziere, Lachaux, Chanussot, and Lample]{roziere2020unsupervised}
Baptiste Roziere, Marie-Anne Lachaux, Lowik Chanussot, and Guillaume Lample.
\newblock Unsupervised translation of programming languages.
\newblock \emph{Advances in neural information processing systems}, 33:\penalty0 20601--20611, 2020.

\bibitem[Roziere et~al.(2021)Roziere, Zhang, Charton, Harman, Synnaeve, and Lample]{roziere2021leveraging}
Baptiste Roziere, Jie~M Zhang, Francois Charton, Mark Harman, Gabriel Synnaeve, and Guillaume Lample.
\newblock Leveraging automated unit tests for unsupervised code translation.
\newblock \emph{arXiv preprint arXiv:2110.06773}, 2021.

\bibitem[Seacord(2013)]{seacord2013secure}
Robert~C Seacord.
\newblock \emph{Secure Coding in C and C++}.
\newblock Addison-Wesley, 2013.

\bibitem[Shetty et~al.(2024)Shetty, Jain, Godbole, Seshia, and Sen]{shetty2024syzygy}
Manish Shetty, Naman Jain, Adwait Godbole, Sanjit~A Seshia, and Koushik Sen.
\newblock Syzygy: Dual code-test c to (safe) rust translation using llms and dynamic analysis.
\newblock \emph{arXiv preprint arXiv:2412.14234}, 2024.

\bibitem[Shiraishi and Shinagawa(2024)]{shiraishi2024context}
Momoko Shiraishi and Takahiro Shinagawa.
\newblock Context-aware code segmentation for c-to-rust translation using large language models.
\newblock \emph{arXiv preprint arXiv:2409.10506}, 2024.

\bibitem[{The Rust Project Developers}(2023)]{rustreference}
{The Rust Project Developers}.
\newblock The rust reference.
\newblock \url{https://doc.rust-lang.org/reference/index.html}, 2023.
\newblock Accessed: December 1, 2024.

\bibitem[Weiss et~al.(2019)Weiss, Gierczak, Patterson, and Ahmed]{weiss2019oxide}
Aaron Weiss, Olek Gierczak, Daniel Patterson, and Amal Ahmed.
\newblock Oxide: The essence of rust.
\newblock \emph{arXiv preprint arXiv:1903.00982}, 2019.

\bibitem[Yang et~al.(2024{\natexlab{a}})Yang, Takashima, Paulsen, Dodds, and Kroening]{yang2024vert}
Aidan~ZH Yang, Yoshiki Takashima, Brandon Paulsen, Josiah Dodds, and Daniel Kroening.
\newblock Vert: Verified equivalent rust transpilation with large language models as few-shot learners.
\newblock \emph{arXiv preprint arXiv:2404.18852}, 2024{\natexlab{a}}.

\bibitem[Yang et~al.(2024{\natexlab{b}})Yang, Liu, Yu, Keung, Li, Liu, Hong, Ma, Jin, and Li]{yang2024exploring}
Zhen Yang, Fang Liu, Zhongxing Yu, Jacky~Wai Keung, Jia Li, Shuo Liu, Yifan Hong, Xiaoxue Ma, Zhi Jin, and Ge~Li.
\newblock Exploring and unleashing the power of large language models in automated code translation.
\newblock \emph{Proceedings of the ACM on Software Engineering}, 1\penalty0 (FSE):\penalty0 1585--1608, 2024{\natexlab{b}}.

\bibitem[Zhang et~al.(2023)Zhang, David, Yu, and Wang]{zhang2023ownership}
Hanliang Zhang, Cristina David, Yijun Yu, and Meng Wang.
\newblock Ownership guided c to rust translation.
\newblock In \emph{International Conference on Computer Aided Verification}, pages 459--482. Springer, 2023.

\bibitem[Zhang et~al.(2024)Zhang, David, Wang, Paulsen, and Kroening]{zhang2024scalable}
Hanliang Zhang, Cristina David, Meng Wang, Brandon Paulsen, and Daniel Kroening.
\newblock Scalable, validated code translation of entire projects using large language models.
\newblock \emph{arXiv preprint arXiv:2412.08035}, 2024.

\end{thebibliography}

\pagebreak
\clearpage
\appendices
\raggedcolumns
\section{Decomposition Algorithm}

\begin{algorithm}[h!]
\newcommand{\FN}[1]{{\javadocblue{\texttt{\footnotesize#1}}}}
\begin{lstlisting}[language=algo, frame=none, style=normal, basicstyle=\footnotesize\ttfamily]
input: G - the AST of the program
input: L - the max length of a translation unit

output: units - a list of pieces of code, each 
        having $\le$ L lines

units := []
$\FN{walk\_subtree}$(G)

fn $\FN{walk\_subtree}$(node):
  for child in node.children:
    match child.type:
      Func => $\FN{visit\_func}$(child),
      Block => $\FN{visit\_block}$(child),
      _ => $\FN{walk\_subtree}$(child)

fn $\FN{visit\_func}$(func):
  // Traverse each function longer than L lines
  if $\FN{line\_count}$(func) <= L: 
    chunks.$\FN{push}$(func)
  else:
    $\FN{walk\_subtree}$(func)

// A block is any entity within {..} parantheses
fn $\FN{visit\_block}$(block):  
  // Walk each subtree longer than L lines
  for stmt in block.statements:
    if $\FN{line\_count}$(stmt) > L:
      $\FN{walk\_subtree}$(stmt)  
  lower, upper := 0, 0 // Maintain two cursors
  while upper $<$ block.statements.$\FN{len}$(): 
    if $\FN{line\_count}$(block.statements[lower..upper]) > L:
      // Adding this statement caused the 
      // length to spill over
      unit := block.statements[lower..upper-1]
      units.$\FN{push}$(unit)
      // Remove this unit from the AST
      G.$\FN{prune\_nodes}$(unit)
      lower = upper
    else:
      upper += 1
  // We have some leftover lines
  if lower $<$ block.statements.$\FN{len}$():
    unit := block.statements[lower..]
    units.$\FN{push}$(unit)
    // Remove this unit from the AST
    G.$\FN{prune\_nodes}$(unit)
\end{lstlisting}
\caption{Decomposing the program into translation units, each having fewer than $L$ lines.}
\label{alg:decomposition}
\end{algorithm}

\section{More Coreutils Results}

In this section, we present some more results on the Coreutils datasets, adding to the results in \Cref{tab:coreutils}. \Cref{tab:coreutils-fromc} contains the results of a variant of our approach that uses direct C to Rust translation. \Cref{tab:cost} shows the number of LLM calls and approximate usage cost for all our benchmarks.
\begin{table}[H]
\caption{Translating CoreUtils to safer Rust using a variant of \tool, with direct C to Rust translation.}
\label{tab:coreutils-fromc}
\begin{minipage}{\linewidth}
\renewcommand{\baselinestretch}{1.25}\selectfont
\centering
\resizebox{0.6\linewidth}{!}{%
\begin{tabular}{lrrrrrr}
\clineB{2-7}{2}
\multicolumn{1}{lV{2}}{\multirow{3}{*}{}} &
  \multicolumn{3}{cV{2}}{Raw Pointer} &
  \multicolumn{3}{cV{2}}{Raw Pointer}\bigstrut[t]\\
\multicolumn{1}{lV{2}}{} &
  \multicolumn{3}{cV{2}}{Declarations ($\nabla$)} &
  \multicolumn{3}{cV{2}}{Deferences  ($\nabla$)} \bigstrut[b]\\ \clineB{2-7}{2}
\multicolumn{1}{lV{2}}{} &
  \multicolumn{1}{rV{2}}{\rotatebox{90}{Before~~}} &
  \multicolumn{1}{rV{2}}{{\rotatebox{90}{After}}} &
  \multicolumn{1}{cV{2}}{$\Delta\%$} &
  \multicolumn{1}{rV{2}}{\rotatebox{90}{Before~~}} &
  \multicolumn{1}{rV{2}}{{\rotatebox{90}{After}}} &
  \multicolumn{1}{cV{2}}{$\Delta\%$} \bigstrut\\ \clineB{2-7}{2}
 &
  \multicolumn{1}{l}{} &
  \multicolumn{1}{l}{} &
  \multicolumn{1}{l}{} &
  \multicolumn{1}{l}{} &
  \multicolumn{1}{l}{} &
  \multicolumn{1}{l}{} \bigstrut\\[-1.33em] \hlineB{2}
\multicolumn{1}{V{2}lV{2}}{split}
& \multicolumn{1}{rV{2}}{252}
& \multicolumn{1}{rV{2}}{232}
& \multicolumn{1}{rV{2}}{\cellcolor{blue!12}8}
& \multicolumn{1}{rV{2}}{656}
& \multicolumn{1}{rV{2}}{628}
& \multicolumn{1}{rV{2}}{\cellcolor{blue!12}4}
\bigstrut\\
\multicolumn{1}{V{2}lV{2}}{pwd}
& \multicolumn{1}{rV{2}}{164}
& \multicolumn{1}{rV{2}}{138}
& \multicolumn{1}{rV{2}}{\cellcolor{blue!12}16}
& \multicolumn{1}{rV{2}}{295}
& \multicolumn{1}{rV{2}}{265}
& \multicolumn{1}{rV{2}}{\cellcolor{blue!12}10}
\bigstrut\\
\multicolumn{1}{V{2}lV{2}}{cat}
& \multicolumn{1}{rV{2}}{192}
& \multicolumn{1}{rV{2}}{163}
& \multicolumn{1}{rV{2}}{\cellcolor{blue!12}15}
& \multicolumn{1}{rV{2}}{317}
& \multicolumn{1}{rV{2}}{289}
& \multicolumn{1}{rV{2}}{\cellcolor{blue!12}9}
\bigstrut\\
\multicolumn{1}{V{2}lV{2}}{truncate}
& \multicolumn{1}{rV{2}}{156}
& \multicolumn{1}{rV{2}}{142}
& \multicolumn{1}{rV{2}}{\cellcolor{blue!12}9}
& \multicolumn{1}{rV{2}}{326}
& \multicolumn{1}{rV{2}}{278}
& \multicolumn{1}{rV{2}}{\cellcolor{blue!12}15}
\bigstrut\\
\multicolumn{1}{V{2}lV{2}}{uniq}
& \multicolumn{1}{rV{2}}{227}
& \multicolumn{1}{rV{2}}{179}
& \multicolumn{1}{rV{2}}{\cellcolor{blue!12}21}
& \multicolumn{1}{rV{2}}{343}
& \multicolumn{1}{rV{2}}{310}
& \multicolumn{1}{rV{2}}{\cellcolor{blue!12}10}
\bigstrut\\
\multicolumn{1}{V{2}lV{2}}{tail}
& \multicolumn{1}{rV{2}}{389}
& \multicolumn{1}{rV{2}}{338}
& \multicolumn{1}{rV{2}}{\cellcolor{blue!12}13}
& \multicolumn{1}{rV{2}}{1092}
& \multicolumn{1}{rV{2}}{1027}
& \multicolumn{1}{rV{2}}{\cellcolor{blue!12}6}
\bigstrut\\
\multicolumn{1}{V{2}lV{2}}{head}
& \multicolumn{1}{rV{2}}{192}
& \multicolumn{1}{rV{2}}{172}
& \multicolumn{1}{rV{2}}{\cellcolor{blue!12}10}
& \multicolumn{1}{rV{2}}{442}
& \multicolumn{1}{rV{2}}{420}
& \multicolumn{1}{rV{2}}{\cellcolor{blue!12}5}
\bigstrut\\
\hlineB{2}
\end{tabular}%
}

\vspace{0.5em}
\subcaption{Number of raw pointer declarations and dereferences.}
\end{minipage}\\
\begin{minipage}{\linewidth}
\renewcommand{\baselinestretch}{1.25}\selectfont
\centering
\resizebox{0.9\linewidth}{!}{%
\begin{tabular}{lrrrrrrrrr}
\clineB{2-10}{2}
\multicolumn{1}{lV{2}}{\multirow{3}{*}{}} &
  \multicolumn{3}{cV{2}}{Unsafe} &
  \multicolumn{3}{cV{2}}{Unsafe} &
  \multicolumn{3}{cV{2}}{Unsafe Call}\bigstrut[t]\\
\multicolumn{1}{lV{2}}{} &
  \multicolumn{3}{cV{2}}{Lines ($\nabla$)} & 
  \multicolumn{3}{cV{2}}{Type Casts ($\nabla$)} &
  \multicolumn{3}{cV{2}}{Expressions  ($\nabla$)} \bigstrut[b]\\ \clineB{2-10}{2}
\multicolumn{1}{lV{2}}{} &
  \multicolumn{1}{rV{2}}{\rotatebox{90}{Before~~}} &
  \multicolumn{1}{rV{2}}{{\rotatebox{90}{After}}} &
  \multicolumn{1}{cV{2}}{$\Delta\%$} &
  \multicolumn{1}{rV{2}}{\rotatebox{90}{Before~~}} &
  \multicolumn{1}{rV{2}}{{\rotatebox{90}{After}}} &
  \multicolumn{1}{cV{2}}{$\Delta\%$} &
  \multicolumn{1}{rV{2}}{\rotatebox{90}{Before~~}} &
  \multicolumn{1}{rV{2}}{{\rotatebox{90}{After}}} &
  \multicolumn{1}{cV{2}}{$\Delta\%$} \bigstrut\\ \clineB{2-10}{2}
& \multicolumn{1}{l}{} &
  \multicolumn{1}{l}{} &
  \multicolumn{1}{l}{} &
  \multicolumn{1}{l}{} &
  \multicolumn{1}{l}{} &
  \multicolumn{1}{l}{} &
  \multicolumn{1}{l}{} &
  \multicolumn{1}{l}{} \bigstrut\\[-1.33em] \hlineB{2}
\multicolumn{1}{V{2}lV{2}}{split}
& \multicolumn{1}{rV{2}}{11324}
& \multicolumn{1}{rV{2}}{10357}
& \multicolumn{1}{rV{2}}{\cellcolor{blue!12}9}
& \multicolumn{1}{rV{2}}{2353}
& \multicolumn{1}{rV{2}}{2326}
& \multicolumn{1}{rV{2}}{\cellcolor{blue!12}1}
& \multicolumn{1}{rV{2}}{11324}
& \multicolumn{1}{rV{2}}{10357}
& \multicolumn{1}{rV{2}}{\cellcolor{blue!12}9}
\bigstrut\\
\multicolumn{1}{V{2}lV{2}}{pwd}
& \multicolumn{1}{rV{2}}{4201}
& \multicolumn{1}{rV{2}}{2921}
& \multicolumn{1}{rV{2}}{\cellcolor{blue!12}30}
& \multicolumn{1}{rV{2}}{875}
& \multicolumn{1}{rV{2}}{851}
& \multicolumn{1}{rV{2}}{\cellcolor{blue!12}3}
& \multicolumn{1}{rV{2}}{4201}
& \multicolumn{1}{rV{2}}{2921}
& \multicolumn{1}{rV{2}}{\cellcolor{blue!12}30}
\bigstrut\\
\multicolumn{1}{V{2}lV{2}}{cat}
& \multicolumn{1}{rV{2}}{5625}
& \multicolumn{1}{rV{2}}{4663}
& \multicolumn{1}{rV{2}}{\cellcolor{blue!12}17}
& \multicolumn{1}{rV{2}}{1038}
& \multicolumn{1}{rV{2}}{950}
& \multicolumn{1}{rV{2}}{\cellcolor{blue!12}8}
& \multicolumn{1}{rV{2}}{5625}
& \multicolumn{1}{rV{2}}{4663}
& \multicolumn{1}{rV{2}}{\cellcolor{blue!12}17}
\bigstrut\\
\multicolumn{1}{V{2}lV{2}}{truncate}
& \multicolumn{1}{rV{2}}{5544}
& \multicolumn{1}{rV{2}}{4581}
& \multicolumn{1}{rV{2}}{\cellcolor{blue!12}17}
& \multicolumn{1}{rV{2}}{1040}
& \multicolumn{1}{rV{2}}{1050}
& \multicolumn{1}{rV{2}}{\cellcolor{blue!12}-1}
& \multicolumn{1}{rV{2}}{5544}
& \multicolumn{1}{rV{2}}{4581}
& \multicolumn{1}{rV{2}}{\cellcolor{blue!12}17}
\bigstrut\\
\multicolumn{1}{V{2}lV{2}}{uniq}
& \multicolumn{1}{rV{2}}{6066}
& \multicolumn{1}{rV{2}}{4986}
& \multicolumn{1}{rV{2}}{\cellcolor{blue!12}18}
& \multicolumn{1}{rV{2}}{1150}
& \multicolumn{1}{rV{2}}{1129}
& \multicolumn{1}{rV{2}}{\cellcolor{blue!12}2}
& \multicolumn{1}{rV{2}}{6066}
& \multicolumn{1}{rV{2}}{4986}
& \multicolumn{1}{rV{2}}{\cellcolor{blue!12}18}
\bigstrut\\
\multicolumn{1}{V{2}lV{2}}{tail}
& \multicolumn{1}{rV{2}}{11663}
& \multicolumn{1}{rV{2}}{9780}
& \multicolumn{1}{rV{2}}{\cellcolor{blue!12}16}
& \multicolumn{1}{rV{2}}{2580}
& \multicolumn{1}{rV{2}}{2485}
& \multicolumn{1}{rV{2}}{\cellcolor{blue!12}4}
& \multicolumn{1}{rV{2}}{11663}
& \multicolumn{1}{rV{2}}{9780}
& \multicolumn{1}{rV{2}}{\cellcolor{blue!12}16}
\bigstrut\\
\multicolumn{1}{V{2}lV{2}}{head}
& \multicolumn{1}{rV{2}}{6245}
& \multicolumn{1}{rV{2}}{5552}
& \multicolumn{1}{rV{2}}{\cellcolor{blue!12}11}
& \multicolumn{1}{rV{2}}{1378}
& \multicolumn{1}{rV{2}}{1383}
& \multicolumn{1}{rV{2}}{\cellcolor{blue!12}-0}
& \multicolumn{1}{rV{2}}{6245}
& \multicolumn{1}{rV{2}}{5552}
& \multicolumn{1}{rV{2}}{\cellcolor{blue!12}11}
\bigstrut\\
\hlineB{2}
\end{tabular}%
}
\vspace{0.5em}
\subcaption{Number of unsafe expressions and declarations.}
\end{minipage}
\end{table}

\begin{table}[ht]
    \centering
    \resizebox{\linewidth}{!} {
    \begin{tabular}{l|r|c|c|c|c}
    \toprule
         & \textbf{Program} & \textbf{\# Fn} & \makecell{\textbf{\# Trans.}\\ \textbf{Units}} & \makecell{\textbf{LLM}\\\textbf{Calls}} & \makecell{\textbf{\textbf{Approx}}\\\textbf{Cost (\$)}} \\
         \bottomrule
\multirow{7}{*}{\rot{90}{\textbf{\textsc{Coreutils}}}} & split & 207 & 292 & 1140 & 0.68 \\
& pwd & 127 & 152 & 527 & 0.31 \\
& cat & 166 & 205 & 636 & 0.38 \\
& truncate & 124 & 170 & 599 & 0.36 \\
& uniq & 167 & 213 & 776 & 0.46 \\
& tail & 266 & 350 & 1224 & 0.73 \\
& head & 153 & 193 & 666 & 0.40 \\
         \midrule
\multirow{10}{*}{\rot{90}{\textbf{\textsc{Laertes}}}} & tulipindicators & 234 & 242 & 1017 & 0.61 \\
& xzoom & 11 & 26 & 101 & 0.06 \\
& genann & 32 & 35 & 124 & 0.07 \\
& optipng & 576 & 941 & 3518 & 2.10 \\
& urlparser & 22 & 25 & 101 & 0.06 \\
& lil & 160 & 184 & 800 & 0.48 \\
& qsort & 3 & 3 & 12 & 0.01 \\
& snudown & 92 & 108 & 488 & 0.29 \\
& grabc & 7 & 7 & 22 & 0.01 \\
& bzip2 & 128 & 251 & 885 & 0.53 \\
        \bottomrule
    \end{tabular}
    }
    \caption{LLM usage cost estimates for translating each benchmark using \tool with \TT{gpt4o-mini}.}
    \label{tab:cost}
\end{table}

\pagebreak

\section{Prompts}

\begin{table}[H]%
\hrule
\begin{lstlisting}[basicstyle=\footnotesize\ttfamily, breaklines=true]
Here is a function:
```rust
unsafe extern "C" fn gettext_quote(
    mut msgid: *const libc::c_char,
    mut s: quoting_style,
) -> *const libc::c_char {
    let mut translation: *const libc::c_char = gettext(msgid);
    let mut locale_code: *const libc::c_char = 0 as *const libc::c_char;
    if translation != msgid {
        return translation;
    }
    locale_code = locale_charset();
    ...
}
```
Here are its call sites
Call site 1:
```rust
right_quote = gettext_quote(
                        b"'\0" as *const u8 as *const libc::c_char,
                        quoting_style,
                    );
```
Call site 2:
```rust
left_quote = gettext_quote(
                        b"`\0" as *const u8 as *const libc::c_char,
                        quoting_style,
                    );
```

The file contains the following imports:
```rust
use ::libc;
```
Convert the function to idiomatic Rust, meaning Rust code that does not make use of features like unsafe, raw pointers, and the C API whenever possible. Do not change the function name.

Follow the following format for your output: Place the function translation between the tags <FUNC> and </FUNC>. If the function signature changed in translation, its callsites will need to be modified as well. Place each callsite translation (in the same order it appears above) between <CALL> and </CALL>. Note that even if the callsite is only a single statement, the translation can be mutiple statements. For example, you may need to declare new variables, or convert between types, either before or after the call. The translation should be such that the surrounding code is not affected by the changes.


Any functions or variables without definitions are defined elsewhere in the code. Do not attempt to redefine them or import them.
If you are using any new functions and you need new imports for those, place them between the tags <IMPORTS> and </IMPORTS>. This should be only *new* imports. Do not include existing imports.
DO NOT include markdown characters like "```" or "```rust" in your translation.
\end{lstlisting}
\hrule
\caption{Translation prompt for a Coreutils function.}
\end{table}

\section{Test Statistics}

\Cref{tab:tests} shows the statistics of the Coreutils test suite that we used to evaluate the correctness of our translations. Each test is a Bash or Perl script that is run against the compiled executable.

\begin{table}[h]
    \centering
    \begin{tabular}{l|r|c|c|c|c|c}
    \toprule
         & \textbf{Program} & \textbf{\# Fn} & \makecell{\rebuttal{}{\textbf{\# Cov.}}\\ \rebuttal{}{\textbf{Fn}}} & \makecell{\textbf{\# Test}\\\textbf{Scripts}} \\
         \bottomrule
\multirow{7}{*}{\rot{90}{\textbf{\textsc{Coreutils}}}} & split & 207 & \rebuttal{}{73} & 12 \\
& pwd & 127 & \rebuttal{}{16} & 2 \\
& cat & 166 & \rebuttal{}{37} & 4 \\
& truncate & 124 & \rebuttal{}{33} & 8 \\
& uniq & 167 & \rebuttal{}{34} & 3 \\
& tail & 266 & \rebuttal{}{76} & 30 \\
& head & 153 & \rebuttal{}{45} & 4 \\
        \bottomrule
    \end{tabular}
    \caption{The statistics of the test suite for the 7 programs in our coreutils benchmark dataset. ``\# Cov. Fn'' is the number of functions covered by tests, and ``\# Test Scripts'' is the number of end-to-end test scripts.}
    \label{tab:tests}
\end{table}

\section{More Qualitative Comparisons}
\label{sec:laertes_comparison}
In this section, we show some more examples of \tool output that will enable us to qualitatively analyze its performance.

\Cref{fig:qualitative} shows an example of the output C2Rust vs Laertes vs \tool. This is taken from \TT{qsort}, one of the Laertes benchmarks. We can see that although both Laertes and \tool eliminate the raw pointer \TT{*mut}, the output of \tool is much more natural and easy to follow.
\begin{figure}[h]
\lstset{escapeinside={<@}{@>}}
 \centering
  \begin{subfigure}{\columnwidth}
    \lstset{xleftmargin=0\columnwidth}
    \begin{lstlisting}[language=Rust, frame=none, style=colouredRust, basicstyle=\footnotesize\ttfamily, numbers=none]
pub unsafe extern "C" fn swap(
        mut a: *mut std::os::raw::c_int,
        mut b: *mut std::os::raw::c_int) {
  let mut t: std::os::raw::c_int = *a;
  *a = *b;
  *b = t;
}
\end{lstlisting}
    \caption{\small{C2Rust output}}
    \end{subfigure} 
    \begin{subfigure}{\columnwidth}
    \lstset{xleftmargin=0\columnwidth}
    \begin{lstlisting}[language=Rust, frame=none, style=colouredRust, basicstyle=\footnotesize\ttfamily, numbers=none]
pub unsafe extern "C" fn swap<'a1, 'a2>(
        mut a: Option<&'a1 mut std::os::raw::c_int>,
        mut b: Option<&'a2 mut std::os::raw::c_int>) {
        
  let mut t: i32 = *(borrow_mut(&mut a)).unwrap();
  *(borrow_mut(&mut a)).unwrap() = 
                    *(borrow_mut(&mut b)).unwrap();
  *(borrow_mut(&mut b)).unwrap() = t;
}
\end{lstlisting}
    \caption{\small{Laertes output.}}
    \label{fig:lifetime-b}
    \end{subfigure} 
    \begin{subfigure}{\columnwidth}
    \lstset{xleftmargin=0\columnwidth}
    \begin{lstlisting}[language=Rust, frame=none, style=colouredRust, basicstyle=\footnotesize\ttfamily, numbers=none]
pub fn swap(a: &mut i32, b: &mut i32) {
  let t = *a;
  *a = *b;
  *b = t;
}
\end{lstlisting}
    \caption{\small{\tool output.}}
    \end{subfigure}
\caption{\small{A qualitative comparison of Laertes and \tool.}}
\label{fig:qualitative}
\end{figure}

\Cref{fig:qualitative2} shows the function \TT{ireallocarray}, which is a common helper function used by all the Coreutils programs. It takes a pointer pointing to potentially uninitialized memory, and allocates memory for it such that the total allocation is equal to a certain specified number of bytes. This is implemented in C with a \TT{void *} pointer. The C2Rust code uses \TT{*mut libc::c\_void}, which is not an idiomatic Rust type. On the other hand, \tool{} correctly handles this by using the type \TT{Option<Vec<u8>>} for \TT{p}. This also allows for the possibility that the memory might be uninitalized, which would correspond to the \TT{None} variant of the \TT{Option} enum.

\begin{figure*}[ht]
\lstset{escapeinside={<@}{@>}}
 \centering
\begin{subfigure}{\columnwidth}
    \lstset{xleftmargin=0\columnwidth}
    \begin{lstlisting}[language=C, frame=none, style=myc, basicstyle=\footnotesize\ttfamily, numbers=none]
/* ireallocarray (ptr, num, size) is like reallocarray (ptr, num, size).
   It returns a non-NULL pointer to num * size bytes of memory.
   Upon failure, it returns NULL with errno set.  */
IALLOC_INLINE void *
ireallocarray (void *p, idx_t n, idx_t s)
{
  if (n <= SIZE_MAX && s <= SIZE_MAX)
    {
      size_t nx = n;
      size_t sx = s;
      if (n == 0 || s == 0)
        nx = sx = 1;
      p = reallocarray (p, nx, sx);
      return p;
    }
  else
    return _gl_alloc_nomem ();
}
\end{lstlisting}
    \caption{\small{C code}}
\end{subfigure}
\begin{subfigure}{\columnwidth}
    \lstset{xleftmargin=0\columnwidth}
    \begin{lstlisting}[language=Rust, frame=none, style=colouredRust, basicstyle=\footnotesize\ttfamily, numbers=none]
pub unsafe extern "C" fn ireallocarray(
    mut p: *mut libc::c_void,
    mut n: idx_t,
    mut s: idx_t,
) -> *mut libc::c_void {
    if n as libc::c_ulong <= 18446744073709551615
            as libc::c_ulong
        && s as libc::c_ulong <= 18446744073709551615
            as libc::c_ulong
    {
        let mut nx: size_t = n as size_t;
        let mut sx: size_t = s as size_t;
        if n == 0 as libc::c_int as libc::c_long
            || s == 0 as libc::c_int as libc::c_long
        {
            sx = 1 as libc::c_int as size_t;
            nx = sx;
        }
        p = reallocarray(p, nx, sx);
        return p;
    } else {
        return _gl_alloc_nomem()
    };
}
\end{lstlisting}
    \caption{\small{C2Rust output.}}
    \label{fig:lifetime-b}
\end{subfigure} 
\begin{subfigure}{\columnwidth}
    \lstset{xleftmargin=0\columnwidth}
    \begin{lstlisting}[language=Rust, frame=none, style=colouredRust, basicstyle=\footnotesize\ttfamily, numbers=none]
pub fn ireallocarray(
    p: Option<Vec<u8>>,
    n: idx_t,
    s: idx_t,
) -> Option<Vec<u8>> {
    if n as usize <= usize::MAX && s as usize <= usize::MAX {
        let mut nx: usize = n as usize;
        let mut sx: usize = s as usize;
        if n == 0 || s == 0 {
            sx = 1;
            nx = sx;
        }
        let mut vec = p.unwrap_or_else(||
                Vec::with_capacity(nx * sx)
            );
        vec.resize(nx * sx, 0);
        return Some(vec);
    } else {
        return None;
    }
}
\end{lstlisting}
    \caption{\small{\tool output.}}
    \end{subfigure}
\caption{\small{A comparison between C, C2Rust, and \tool, for an example involving memory allocation. \tool appropriately handles the allocation using \TT{Option<Vec<u8>>} in Rust.}}
\label{fig:qualitative2}
\end{figure*}

\end{document}